\documentclass[%
 reprint,
 amsmath,amssymb,
 aps,
]{revtex4-1}

\usepackage{graphicx}

\usepackage{dcolumn}
\usepackage{bm}
\usepackage{multirow} 

\usepackage{placeins}

\usepackage{colortbl}
\usepackage{url}
\usepackage{nicefrac}
\usepackage{units}    
\usepackage{hyperref}

\PassOptionsToPackage{table}{xcolor}

\usepackage{tikz}
\usepackage{xcolor}  

\usepackage{algorithm}
\usepackage{booktabs}
\usepackage{amssymb}  
\graphicspath{{Figures/}} 

\renewcommand{\vec}[1]{\boldsymbol{#1}}

\usepackage{caption}
\usepackage{marginnote}
\usepackage[normalem]{ulem}

\begin{document}

\preprint{PRX LIFE}

\title{Gait Transitions in Load-Pulling Quadrupeds: \texorpdfstring{\\}{ }
Insights from Sled Dogs and a Minimal SLIP Model}

\author{Jiayu Ding\,}%
\thanks{These authors contributed equally to this work.}%
\affiliation{College of Engineering and Computer Science, Syracuse University, Syracuse, NY 13244, USA}

\author{Benjamin Seleb\,}%
\thanks{These authors contributed equally to this work.}%
\affiliation{Interdisciplinary Graduate Program in Quantitative Biosciences, Georgia Institute of Technology, Atlanta, GA 30332, USA}

\author{Heather J. Huson\,}
\affiliation{Department of Animal Science, Cornell University, Ithaca, NY 14853, USA}

\author{Saad Bhamla\,}%
\thanks{These authors are corresponding authors of this work.}%
\affiliation{School of Chemical and Biomolecular Engineering, Georgia Institute of Technology, Atlanta, GA 30332, USA}

\author{Zhenyu Gan\,}%
\thanks{These authors are corresponding authors of this work.}%
\affiliation{College of Engineering and Computer Science, \ Syracuse University, Syracuse, NY 13244, USA}

\date{\today}

\begin{abstract}

Quadrupedal animals employ diverse galloping strategies to optimize speed, stability, and energy efficiency. However, the biomechanical mechanisms that enable adaptive gait transitions during high-speed locomotion under load remain poorly understood. 
In this study, we present new empirical and modeling insights into the biomechanics of load-pulling quadrupeds, using sprint sled dogs as a model system. 
High-speed video and force recordings reveal that sled dogs often switch between rotary and transverse galloping gaits within just a few strides and with minimal changes in speed and stride duration, suggestive of of locomotor multistability during high-speed load-pulling.
To investigate the mechanical basis of these transitions, a physics-based quadrupedal Spring-Loaded Inverted Pendulum model with hybrid dynamics and prescribed footfall sequences to reproduce the asymmetric galloping patterns observed in racing sled dogs.
Through trajectory optimization, we replicate experimentally observed gait sequences and identify swing-leg stiffness modulation as a key control mechanism for inducing transitions.
This work provides a much-needed biomechanical perspective on high-speed animal draft and establishes a modeling framework for studying locomotion in pulling quadrupeds, with implications for both biological understanding and the design of adaptive legged systems.

\begin{description}
\item[DOI]
TBD.

\item[Keywords]
\ \ Gait transitions in quadrupeds, \ \ Load-pulling locomotion dynamics, \ \ Sled dogs, \ \ \\
Spring-loaded 
inverted pendulum modeling

\end{description}
\end{abstract}

\maketitle

\section{Introduction}

\begin{figure*}[t]
    \centering
    \includegraphics[width=0.95\textwidth]{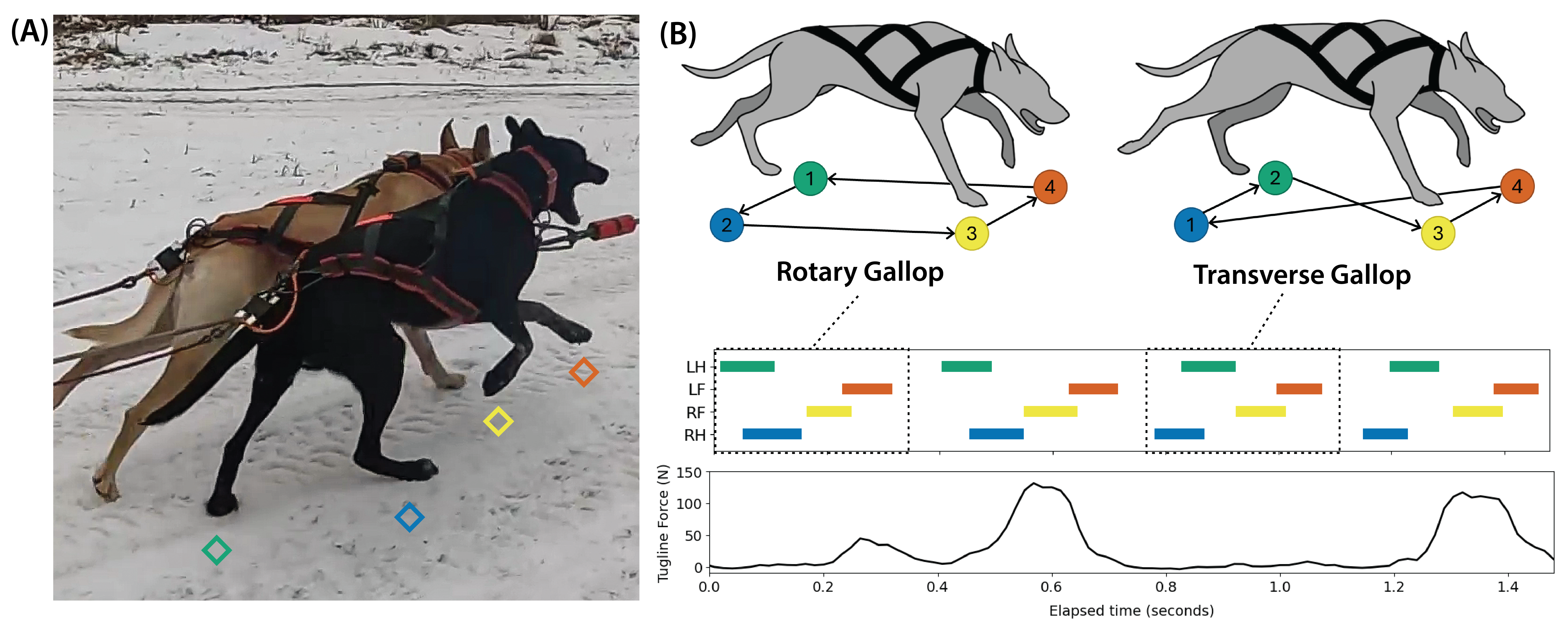}
    \caption{Sled dog gait analysis. (A) A video frame showing sled dogs in motion, with colored markers used to indicate each foot. Similar markers were used to identify stance phases during the stride, labeled manually. 
    (B) A segment of reconstructed footfall data and recorded tugline force. The stance-stride plot illustrates footfall timing, with the colored bars representing the stance durations for each limb. Cartoons depict the ordering of the legs in the corresponding rotary and transverse gait patterns present. The synchronized force recording measures the dynamic loading of the tugline throughout the sled dogs' gait.}
    \label{fig:intro}
\end{figure*}

Coordinated limb movement is fundamental to effective animal locomotion~\cite{biewener2003animal,dickinson2000animal}. Among terrestrial mammals, quadrupedal galloping is one of the most mechanically dynamic and energetically demanding gait patterns. Species such as dogs, horses, and cheetahs adopt distinct galloping strategies—particularly \textit{rotary} and \textit{transverse} variants—to optimize performance across different speeds and terrains~\cite{biancardi2012biomechanical,Hildebrand1977}. 
Traditionally, galloping has been viewed as a stereotyped high-speed gait, with transitions primarily linked to locomotor speed and related energetics~\cite{Hoyt1981gaitenergetic}. More recent work in biomechanics and legged robotics suggests that gait preferences and transitions are not governed by speed alone, but can instead emerge from feedback-driven responses to stability demands, external loading, and terrain-dependent perturbations~\cite{shafiee2024viability,Farley1991trigger,owaki2017quadruped,zhang2024learning,wilshin2017longitudinal}. Lead limb switching within asymmetrical gaits has also been proposed as a wear-leveling strategy to redistribute limb-specific demand and mitigate fatigue~\cite{walter2007ground}.
These perspectives motivate a reinterpretation of galloping (and other gaits) through the lens of \textit{locomotor multistability}, in which multiple gait solutions can coexist under nearly identical conditions and be selectively expressed in response to perturbations or control objectives~\cite{wilshin2020dog}.
For instance, modeling studies have shown that even small variations in limb compliance or body symmetry can give rise to distinct galloping modes within the same speed range~\cite{Ding2024RALSymmetry,alqaham202516}. 
Despite growing interest in context-dependent gait selection and switching, stride-resolved studies of high-speed galloping under sustained external load (e.g., harness-mediated pulling) remain rare.
Sled dogs offer a particularly compelling and under-studied example of this phenomenon: they sustain high-speed galloping while pulling a sled, requiring coordination not only for self-propulsion but also to generate forward force along a connecting rope known as the \textit{tugline}. This added load introduces unique biomechanical constraints, as the resulting tension likely affects limb loading and possibly interlimb timing.

Although our study focuses on biological quadrupeds, the mechanisms uncovered here are also relevant for the design and control of agile, load-pulling quadrupedal robots.
Despite growing interest in quadrupedal locomotion, particularly in working contexts, the dynamics of load-pulling quadrupeds remain poorly understood. A long history of animal draft in agriculture, transportation, and sport has generated limited biomechanical literature on load-pulling quadrupeds, especially under dynamic conditions. Most existing work has focused on practical questions in agricultural settings, emphasizing work output or harness efficiency during heavy-load, static-pulling scenarios~\cite{starkey1989animal,starkey1989harnessing,lawrence1993experimental,Rooney1985,Bukhari2023}. Recent robotics research has begun to model dynamic quadruped-load systems~\cite{RobotLoadPathPlaning2024}, but such models typically operate at the path-planning level and do not capture limb-level dynamics or the energetic coupling between the quadruped and its load. To the best of our knowledge, existing dynamical models of quadruped–load systems do not simultaneously capture limb–load coupling and stride-level coordination in a way that reproduces the experimentally observed gait transitions.

In this study, we contribute new empirical and modeling insights to the study of load-pulling quadrupeds. Using high-speed video recordings of racing sled dogs, we reconstructed footfall sequences and tugline forces during high-speed locomotion. Our analysis reveals that sprint sled dogs do not rely on a single fixed gallop. Instead, they frequently switch between rotary and transverse galloping patterns, often within a few strides and without large changes in forward speed, consistent with a multistable gait repertoire.

We build on the Spring-Loaded Inverted Pendulum (SLIP) framework~\cite{geyer2006spring,full1999templates,fukuhara2018spontaneous}, a reduced-order model widely used to study locomotor dynamics, and extend it to quadrupedal systems that incorporate limb asymmetry, external load pulling, and predefined footfall sequences. While prior models have explored gait energetics~\cite{alqaham2024sixteen} and bipedal transitions~\cite{DingJerboa2022}, our formulation captures the full-body dynamics and interaction forces between a quadruped and a load. In contrast to neuromechanical simulations that require detailed anatomical modeling~\cite{geyer2010muscle}, our model achieves generalizability with fewer parameters while remaining grounded in physical dynamics.
To our knowledge, this is the first dynamical model capable of reproducing not only the detailed stride mechanics but also the spontaneous gait transitions observed in load-bearing quadrupeds. By embedding empirical gait sequences and optimizing against experimental data, we show that these transitions can be generated primarily through targeted modulation of limb torsional stiffness, with hind-limb stiffness acting as a dominant control-like parameter and forelimb stiffness enabling changes in leading-leg sequence. This identifies differential limb stiffness as a simple, mechanically grounded strategy for organizing stride-to-stride transitions in quadruped–load systems and delineates the region of parameter space in which such transitions are feasible.

\section{Gait Transitions in Sprint Sled Dogs}

\subsection{Gait Pattern Identification}
\label{subsec:ID}

Fieldwork was conducted in collaboration with a professional sprint-racing sled dog kennel, with data collection efforts integrated into their regular training routine. Sprint-racing is a fast-paced style of dogsled racing, in which dogs typically maintain galloping speeds of 25 to 40 km per hour over distances less than 50 km \cite{thorsrud2021description}. Training runs took place on dirt forest roads or shallow snow, with teams pulling a motorized ATV carrying two human passengers. We restricted analysis to straight, steady-running segments. Specifically, we excluded intervals containing starts/stops, overt turns, or obvious driver-induced speed changes. Throughout, the driver aimed to maintain race-like speeds, but speed and acceleration were not strictly fixed.

Precise measurement of limb-specific stance and stride timing typically relies on controlled experiments using high-speed video, with either manual annotation \cite{Hilderbrand1989Quadrupedal,sandberg2020review} or automated labeling methods \cite{sheppard2022stride}. However, replicating such methods in the field, especially with free-running sled dogs, is logistically challenging. To overcome this, we mounted an action camera (GoPro, 60 fps +) on an extended pole held from the training rig. This setup enabled us to capture targeted side-view footage of two individuals—referred to here as Individual 1 and Individual 2—during select training runs. From this footage, we manually annotated stance and stride durations for each limb (Fig.~\ref{fig:intro}A).

These annotations were used to classify gait patterns based on the unique sequence of limb-ground interactions. Following Hildebrand's convention~\cite{Hildebrand1977}, \textit{a footfall diagram} can be used to depict contact durations as solid bars and aerial phases as gaps. A short selection of footfalls, presented this way (Fig.~\ref{fig:intro}B), illustrates the frequent switching between transverse and rotary gallops during high-speed pulling.
In \textit{transverse gallop}, the hind legs strike in sequence, followed by the contralateral and then ipsilateral forelegs. In \textit{rotary gallop}, the placement of the second hind foot is followed by the ipsilateral forefoot, creating a seemingly rotating sequence of footfalls around the body \cite{biancardi2012biomechanical}.
All four permutations of these galloping patterns are illustrated in Fig.~\ref{fig: galloping def 2}.

Our analysis also benefits from data collected in parallel with video using custom data loggers integrated into the dogs’ harnesses. These sensors were clipped onto a standard `X-back' harness near the dog's withers, 
and measured tugline force (62+ Hz) and tri-axial acceleration (120+ Hz) throughout each stride. After experiments were completed, data was transferred from the data loggers to a computer for post-hoc analysis. For this paper, only the force data was analyzed. 
For a more detailed overview of the sensors used, including design files and firmware, see the Data and materials availability section. The GoPro camera recorded GPS data with absolute (epoch) timestamps in the video file metadata, and the data logger used an onboard GPS module to timestamp each sensor sample (force and acceleration) on the same epoch time base. During analysis, we synchronized the video frames with the logger data by aligning these GPS timestamps.Fig.~\ref{fig:intro}B shows this tension over a short period of time. Notably, the measured force exhibits stride-synchronized oscillations, with peaks that often—but not always—coincide with the touch-down of the trailing forelimb. Load cells were calibrated between runs using fixed weights to correct for mechanical drift or deformation.

Dog stride frequency was computed stride by stride as the time between consecutive touchdowns of the trailing hind limb, which approximates the full galloping cycle. While the animal-borne sensors had GPS, these were used solely for temporal synchronization. Forward speed was instead estimated from a separate GPS receiver mounted on the ATV. Latitude and longitude were converted to planar coordinates, and instantaneous ground speed was computed as the magnitude of the central-difference displacement divided by the corresponding time interval and then smoothed with a 5-s centered moving-average filter. Because the dogs were continuously tethered to the ATV along straight, approximately level road segments, we treated ATV speed as the forward speed of the team. The resulting time series of stride frequency, forward speed, and timing of identified gait transitions are shown in Fig.~\ref{fig:gait_statistics}.

\subsection{Gait Pattern Definitions}
\label{subsec:gallop}

\begin{figure*}[thb]
\centering
\includegraphics[width=0.95\textwidth]{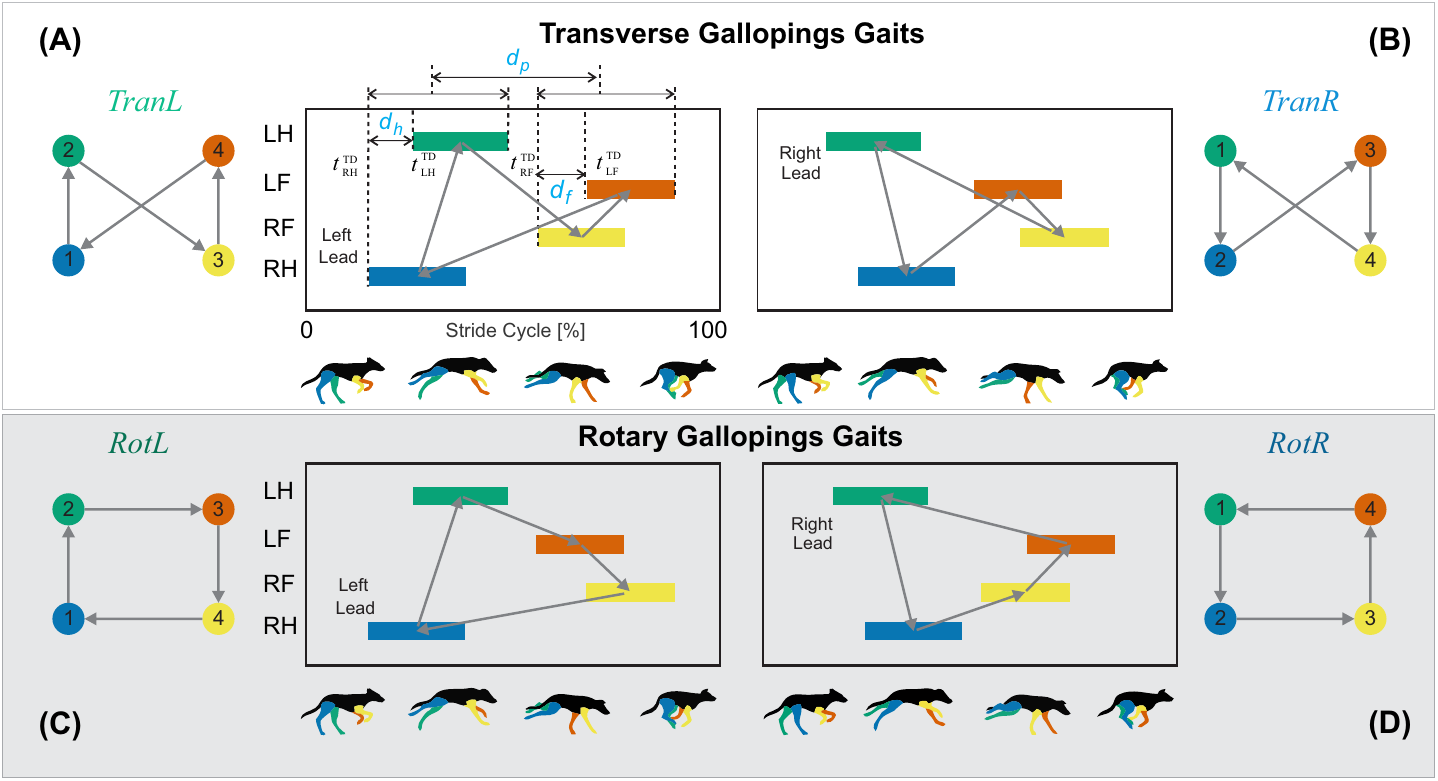}
\caption{Quadrupedal footfall patterns following Hildebrand's convention~\cite{Hildebrand1977} showing (A) Transverse galloping, left leading (TranL), (B) Transverse galloping, right leading (TranR), (C) Rotary galloping, left leading (RotL), (D) Rotary galloping, right leading (RotR). The horizontal axis represents normalized stride time. Colored bars denote ground contact phases for each leg: green (LH), red (LF), yellow (RF), and blue (RH). Each row represents a distinct galloping gait, with white backgrounds for transverse sequences and grey for rotary. Touchdown ($t_i^{\scriptstyle\text{TD}}$) and liftoff ($t_i^{\scriptstyle\text{LO}}$) define the stance period. Phase lags are annotated for fore-hind ($d_p$), hindlimb ($d_h$), and forelimb ($d_f$) timing. Below each bar plot, cartoons depict limb configurations, and adjacent dot plots show footfall sequences, with green and blue font labels for left- and right-leading gaits.}
\label{fig: galloping def 2}
\end{figure*}

Each galloping gait is defined by the touchdown (TD) and liftoff (LO) events of all four limbs over a single stride cycle. We assume that each leg undergoes exactly one stance and one flight phase per stride. Limb labels are assigned as follows: left hind (LH), left fore (LF), right fore (RF), and right hind (RH).

To standardize comparisons across gait types, we define an extended flight phase between the LO of the trailing hindlimb and the TD of the leading forelimb, during which the torso is airborne. The start of each stride is anchored to the midpoint of this aerial phase, corresponding to the moment of peak torso height. This convention follows biomechanical interpretations of galloping dynamics and ensures a consistent phase reference for stride analysis~\cite{Ding2024RALSymmetry, DingJerboa2022}.
Each stride is encoded by the timing vector:
\begin{equation}
    \vec{t} = [t^{\scriptstyle\text{TD}}_{\scriptstyle\text{LH}},\ t^{\scriptstyle\text{LO}}_{\scriptstyle\text{LH}},\ t^{\scriptstyle\text{TD}}_{\scriptstyle\text{LF}},\ t^{\scriptstyle\text{LO}}_{\scriptstyle\text{LF}},\ t^{\scriptstyle\text{TD}}_{\scriptstyle\text{RF}},\ t^{\scriptstyle\text{LO}}_{\scriptstyle\text{RF}},\ t^{\scriptstyle\text{TD}}_{\scriptstyle\text{RH}},\ t^{\scriptstyle\text{LO}}_{\scriptstyle\text{RH}}].
    \label{eq:timing}
\end{equation}
Each element $ t_i $, for $ i\in\{\small\text{LH}, \small\text{LF}, \small\text{RF}, \small\text{RH}\} $, is normalized to the stride interval $ [0,1) $ w.r.t. the total stride time $T$ (in seconds). The timing vector is organized by anatomical order rather than temporal sequence, ensuring consistent limb labeling across gait types.

Using the timing vector, we define several normalized metrics to quantify key temporal relationships:
\newcommand{\wrap}[1]{\left( #1 + 0.5 \right) \bmod 1 - 0.5}
\begin{itemize}
    \item \textbf{Duty factor:} $ d_t \in [0,1) $ is the stance duration for a given leg:
    \[ 
    d_t := (t^{\scriptstyle\text{LO}}_i - t^{\scriptstyle\text{TD}}_i) \bmod 1.
    \]
    The modulo handles wrapping when liftoff occurs before touchdown in normalized stride time.

    \item \textbf{Forelimb phase lag:} $ d_f \in (-0.5, 0.5] $ measures the TD timing difference between left and right forelimbs:
    \[
    d_f := \left[ \left( t^{\scriptstyle\text{TD}}_{\scriptstyle\text{LF}} - t^{\scriptstyle\text{TD}}_{\scriptstyle\text{RF}} + 0.5 \right) \bmod 1 \right] - 0.5.
    \]

    \item \textbf{Hindlimb phase lag:} $ d_h \in (-0.5, 0.5] $ is defined similarly:
    \[
    d_h := \left[ \left( t^{\scriptstyle\text{TD}}_{\text{LH}} - t^{\scriptstyle \text{TD}}_{\scriptstyle \text{RH}} + 0.5 \right) \bmod 1 \right] - 0.5.
    \]

    \item \textbf{Fore-hind phase lag:} $ d_p \in [-0.5, 0.5) $ compares average mid-stance timing between hindlimbs and forelimbs:
    \[
    d_p := \left[ \frac{1}{2} \left( \bar{t}_{\scriptstyle \text{LH}} + \bar{t}_{\scriptstyle \text{RH}} - \bar{t}_{\scriptstyle \text{LF}} - \bar{t}_{\scriptstyle \text{RF}} \right) + 0.5 \right] \bmod 1 - 0.5,
    \]
    where $ \bar{t}_i := \frac{1}{2}(t^{\scriptstyle\text{TD}}_i + t^{\scriptstyle\text{LO}}_i) $ denotes the mid-stance time of leg $ i $.
\end{itemize}

With the timing variables and phase lag metrics defined above, we now define the space of asymmetric galloping gaits as those in which both forelimb and hindlimb phase lags are neither zero nor equal to $ 0.5 $:
\begin{equation}
    G_e = \left\{ \vec{t} \,\middle|\, d_f ,\ d_h \notin \{0, 0.5\} \right\}.
\end{equation}
Phase lags of zero correspond to bounding or half-bounding gaits, while values of $ 0.5 $ indicate symmetrical gaits such as pacing or trotting, following Hildebrand's classification~\cite{Hildebrand1977}.
Within this space, each canonical galloping gait $ G_e $, for $ e \in \{\textit{TranL}, \textit{TranR}, \textit{RotL}, \textit{RotR} \} $, is uniquely determined by the sign pair $ (d_f, d_h) $:
\begin{itemize}
  \item \textit{TranL} (transverse, left-leading): $d_f > 0,\ d_h > 0$
  \item \textit{TranR} (transverse, right-leading): $d_f < 0,\ d_h < 0$
  \item \textit{RotL} (rotary, left-leading): $d_f > 0,\ d_h < 0$
  \item \textit{RotR} (rotary, right-leading): $d_f < 0,\ d_h > 0$
\end{itemize}

These gait labels are visually depicted in Fig.~\ref{fig: galloping def 2}. Transverse gallops appear in the top row (white background), and rotary gallops in the bottom row (grey background). Colored bars indicate stance phases; schematic illustrations below each plot highlight limb movement; and dot plots emphasize the sequence of footfalls and leading limb.
Gait transitions correspond to sign changes in either $d_f$ or $d_h$. For instance, a transition from \textit{TranL} to \textit{RotL} involves a reversal in hindlimb phasing (change in $d_h$), while a switch from \textit{RotL} to \textit{RotR} reflects a forelimb phase reversal (change in $d_f$). Up to 12 distinct transitions can be theoretically defined among the four canonical galloping gaits based on the sign combinations of phase lags. These transitions form the basis for analyzing discrete gait-switching behaviors observed in sled dogs.

\section{Hybrid SLIP-Based Modeling of Load-Pulling Galloping Gaits}
\label{sec:SLIPModel}
This section presents a physics-based modeling framework designed to reproduce and analyze galloping gait patterns in sled dogs pulling a load. We begin by introducing a planar quadrupedal SLIP model extended with a tugline-coupled load, followed by a hybrid dynamics formulation that captures discrete gait transitions through leg touchdown and liftoff events. We then describe an optimization-based approach for identifying periodic solutions that closely match experimentally observed footfall timings and tugline forces.

\subsection{Quadrupedal SLIP Model with Load}
\label{sec:SLIPModel}

We extend a planar SLIP model to capture the dynamics of sled dogs pulling a load at high speed. The base model, adapted from our previous work~\cite{GanDynamicSimilarity,Ding2024RALSymmetry}, consists of a rigid torso supported by four massless spring legs and augmented with a discrete load mass connected via a unilateral spring element that mimics the intermittent tension of a tugline.

\begin{figure}[tbh]
    \centering
    \includegraphics[width=1\columnwidth]{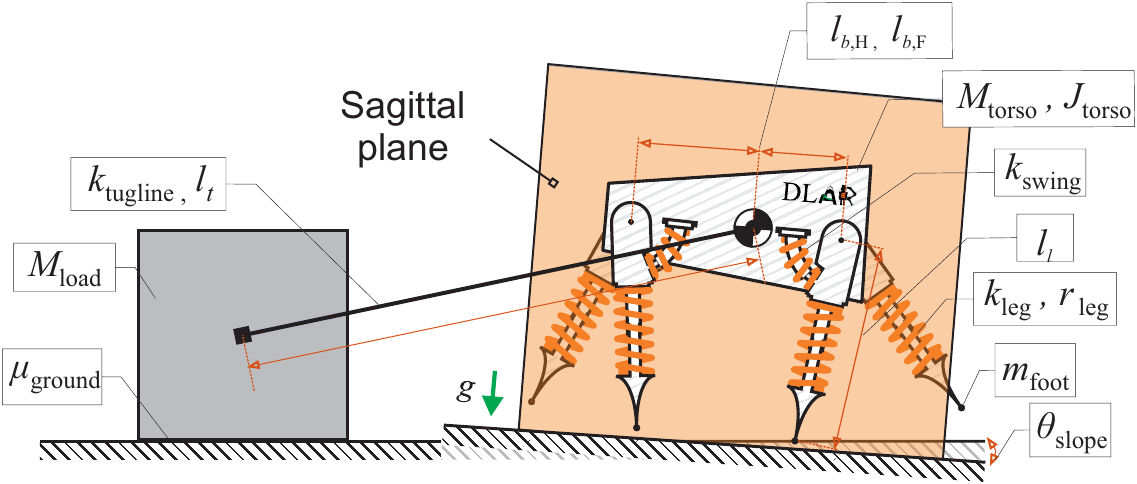}
    \caption{The quadrupedal SLIP model used in this work, with legs connected to the torso through both linear and torsional springs. The model is constrained to planar sagittal motion. Each leg is passively actuated and exhibits distinct swing and stance dynamics. The pulling load is attached via a unilateral spring element modeling the tugline.}
    \label{fig:model}
\end{figure}

The quadruped consists of a rigid body of mass $M_{\text{torso}}$ and moment of inertia $J_{\text{torso}}$, supported by four limbs anchored at fixed hip and shoulder joints. 
The system evolves in the sagittal plane and is described by eight generalized coordinates: the torso position $(q_x, q_z)$, pitch angle $q_\theta$, four leg swing angles $q_{\alpha,i}$ for $ i\in\{\small\text{LH}, \small\text{LF}, \small\text{RF}, \small\text{RH}\} $, and the relative horizontal displacement of the load $q_{\text{load}}$. Because the model is constrained to sagittal-plane motion and does not include limb-specific costs or fatigue dynamics, it is effectively symmetric with respect to left- versus right-leading solutions.

The model forms a hybrid dynamical system, with stance and flight phases governed by distinct sets of equations. 
The hind and fore legs are mounted at distances $l_{b,H}$ and $l_{b,F}$ from the torso Center of Mass (CoM), respectively. 
In stance, ground contact imposes holonomic constraints and generates ground reaction forces (GRFs); in flight, limbs follow unconstrained passive dynamics driven by torsional stiffness.
Each leg behaves as a linear spring with stiffness $k_{\text{leg}}$ during stance, storing and releasing energy to propel the body forward. Legs are massless but terminate in point feet with small masses $m_{\text{foot}}$ to define discrete ground contact events \cite{GanAllCommonBipedal}. A fore–hind stiffness ratio $r_{\text{leg}}$ is included to model anatomical asymmetry between limb pairs.
During flight, each leg swings passively under a torsional spring of stiffness $k_{\text{swing}}$, resulting in approximately sinusoidal motion. Biologically, $k_{\text{swing}}$ represents an effective hip/shoulder swing impedance modulated by muscle activation, while $k_{\text{leg}}$ represents an effective stance-phase limb stiffness arising from musculoskeletal compliance and activation.

To represent the external load, we introduce a point mass $M_{\text{load}}$ connected to the torso by a unilateral, massless spring (the tugline) with stiffness $k_{\text{tug}}$ and resting length $l_t$. The tugline exerts force only when stretched beyond its rest length, producing discontinuous coupling between the quadruped and the load. When slack, the load follows its inertial trajectory; when taut, it exerts a horizontal restoring force that influences both the load and torso dynamics.
To maintain approximately steady galloping in the presence of energy loss (e.g., due to sliding friction at the load), similar to the inverted pendulum model with rigid legs \cite{kuo2005energetic}, we adopt a ``tilted-gravity'' formulation. Specifically, we tilt the terrain by a small slope angle $\theta_{\text{slope}}$, so that gravity acquires a constant component along the direction of motion, and we model Coulomb friction at the load--ground interface with coefficient $\mu_{\text{ground}}$. The resulting downslope component of gravity is mathematically equivalent to a constant horizontal pulling force and can be interpreted as a minimal lumped representation of the net directional work exerted by the dog through the tugline; this effective pull compensates frictional and other dissipative losses and thereby enables sustained periodic motion without introducing additional limb-level actuation parameters.
All parameters are summarized in Table~\ref{tab:params} and are normalized by the total system mass $m$, gravitational acceleration $g$, and resting leg length $l_0$.
The initial values of these parameters are also reported in Table~\ref{tab:params}, where the quadruped model parameters were inherited from the model in \cite{Ding2024RALSymmetry}, which was identified from animal data.

\begin{table}[t]
\centering
\scriptsize
\caption{Key model parameters used in the quadrupedal SLIP model with load. All quantities are normalized using total mass $m$, gravity $g$, and resting leg length $l_0$~\cite{GanAllCommonBipedal}.}
\label{tab:params}
\begin{tabular}{c|c|c|c}
\toprule
\textbf{Symbol} & \textbf{Description} & \textbf{Units}  &\textbf{Initial Value} \\
\midrule
$M_{\rm torso}$       & Torso mass & $m$  &1\\
$m_{\rm foot}$        & Mass of each foot & $m$ &0\\
$J_{\rm torso}$       & Torso pitching inertia & $m l_0^2$ &1.1\\
$k_{\rm leg}$         & Leg spring stiffness & $m g / l_0$ &10\\
$r_{\rm leg}$         & Stiffness ratio between hind/fore legs  & $\cdot$  &1\\
$ k_{\mathrm{swing},i}^n $       & Torsional stiffness of swing‐leg spring & $m g l_0$ &20\\
$l_l$                 & Resting leg length & $l_0$  &1\\
$l_{b,H}$, $l_{b,F}$  & Distance from CoM to hip/shoulder & $l_0$ &0.5\\
\hline
$M_{\rm load}$        & Load mass & $m$ &0.1\\
$k_{\rm tug}$         & Tugline (unilateral spring) stiffness & $m g / l_0$ &10\\
$l_t$                 & Tugline resting length & $l_0$ &2\\
$\mu_{\text{ground}}$ & Friction coefficient of ground contact &$\cdot$ &0.1\\
$\theta_{\text{slope}}$ & Slope angle of running surface & $\cdot$ &0.1\\
\bottomrule
\end{tabular}
\end{table}

To enable gait transitions and capture limb-specific asymmetries observed in natural locomotion, each leg is assigned an independent swing-leg stiffness parameter $ k_{\rm swing} $, which may vary between consecutive strides. For an $ N $-stride sequence, the stiffness set is denoted as $ \vec{k}_{\rm swing} = \{ k_{\mathrm{swing},i}^n \} $, where $ i \in \{\small\text{LH}, \small\text{LF}, \small\text{RF}, \small\text{RH}\} $ and $ n = 1{:}N $. This formulation supports stride-wise modulation of limb stiffness and enables replication of experimentally observed gait transitions driven by localized changes in limb dynamics.

\subsection{Hybrid Dynamics of the Quad-load System}

The quadrupedal SLIP model with tugline evolves as a hybrid dynamical system, where continuous dynamics change with discrete contact states. The configuration vector is defined as 
$\vec{q} = [q_x,\ q_z,\ q_{\theta},\ \vec{q}_{\alpha}^\top,\ q_{\text{load}}]^\top \in \mathbb{R}^8,$
and the continuous dynamics are governed by the Euler-Lagrange equations:
\begin{align}
\vec{M}(\vec{q}) \ddot{\vec{q}} + \vec{C}(\vec{q}, \dot{\vec{q}})\dot{\vec{q}} + \vec{G}(\vec{q}) = \sum_i \vec{J}_i^\top(\vec{q}) \boldsymbol{\lambda}_i,
\label{eq:EOM}
\end{align}
where $ \vec{M}(\vec{q}) \in \mathbb{R}^{8 \times 8} $ is the mass-inertia matrix, $ \vec{C}(\vec{q}, \dot{\vec{q}})\dot{\vec{q}} $ captures Coriolis and centrifugal terms, and $ \vec{G}(\vec{q}) $ encodes gravitational forces. Each contact force $ \boldsymbol{\lambda}_i \in \mathbb{R}^2 $ is transmitted through a Jacobian $ \vec{J}_i(\vec{q}) \in \mathbb{R}^{2 \times 8} $, which maps GRFs into generalized coordinates.

We assume that each leg undergoes exactly one touchdown and one liftoff per stride to avoid Zeno behavior~\cite{pace2017piecewise, ames2005sufficient}. As a result, the eight timing variables $ t_i $ in \eqref{eq:timing} partition the hybrid gait cycle into domains $ j = 1{:}9 $, each corresponding to a unique combination of ground-contacting legs. Within each domain, the associated holonomic constraints are enforced using stacked Jacobians $ \vec{J}_j $ \cite{alqaham2024sixteen}, yielding the following differential-algebraic formulation:
\begin{equation}
\label{eq:DAE} 
\mathcal{F}_{j} :
\begin{bmatrix}
\vec{M}(\vec{q}) & -\vec{J}_j^\intercal(\vec{q}) \\
\vec{J}_j(\vec{q}) & \vec{0}
\end{bmatrix}
\begin{bmatrix}
\ddot{\vec{q}} \\
\vec{\lambda}_j
\end{bmatrix}
=
\begin{bmatrix}
-\vec{C}(\vec{q}, \dot{\vec{q}})\dot{\vec{q}} - \vec{G}(\vec{q}) \\
-\dot{\vec{J}}_j(\vec{q}, \dot{\vec{q}})\dot{\vec{q}}
\end{bmatrix},
\end{equation}
where $ \vec{\lambda}_j $ includes all GRFs in domain $ \mathcal{F}_j $. Each of the $ 2^4 = 16 $ possible contact combinations defines a continuous domain in the hybrid system. 

Touchdown and liftoff events for each leg $ i $ define the switching surfaces:
\begin{equation}
\begin{split}
\mathcal{C}^{\scriptstyle\text{TD}}_i &= \left\{ (\vec{q}, \dot{\vec{q}}) \in \mathcal{TQ} \,\middle|\, l_i(\vec{q}) = l_l,\ \dot{l}_i(\vec{q}) < 0 \right\}, \\
\mathcal{C}^{\scriptstyle\text{LO}}_i &= \left\{ (\vec{q}, \dot{\vec{q}}) \in \mathcal{TQ} \,\middle|\, l_i(\vec{q}) = l_l,\ \dot{l}_i(\vec{q}) > 0 \right\},
\end{split}
\end{equation}
where $ l_i(\vec{q}) $ is the actual leg length and $ l_l $ is the rest length. 
The hybrid model is written as:
\begin{equation}
\label{eq:HM}
\mathcal{H} :
\left\{
\begin{array}{ll}
\mathcal{F}_j, & (\vec{q}, \dot{\vec{q}}) \notin \mathcal{C}_j(t), \\
\dot{\vec{q}}^+ = \Delta_{\mathcal{F}_j \rightarrow \mathcal{F}_{j+1}} \dot{\vec{q}}^-, & (\vec{q}, \dot{\vec{q}}) \in \mathcal{C}_j(t),
\end{array}
\right.
\end{equation}
where $ \Delta $ denotes the impact map that instantaneously resets velocities at transitions. Further implementation details are provided in~\cite{Ding2024RALSymmetry, westervelt2018feedback, alqaham2024sixteen} %

\subsection{Trajectory Optimization for Gait Replication}
\label{sec:opti}

To reproduce periodic gaits and gait transitions observed in sled dogs, we formulate a trajectory optimization problem that identifies feasible motion trajectories of the hybrid SLIP-load model while minimizing a composite cost. The objective function integrates physical consistency, alignment with experimentally recorded footfall timings and tugline forces, and stride duration fidelity.

The optimization is performed over a reduced set of decision variables, including the initial state $ (\vec{q}_0, \dot{\vec{q}}_0) $ and model parameters $ \vec{p} $, which consist of all entries listed in Table~\ref{tab:params}. Among these, only the torso center-of-mass locations $ l_{b,H} $ and $ l_{b,F} $ are held fixed to preserve anatomical consistency.
Let $ N $ denote the number of strides and the objective is expressed as: 
\begin{align}
\min_{\substack{
    \vec{q}_0,\ \dot{\vec{q}}_0, \ \vec{p}\\
}} &  \ \ \sum_{n=1}^{N} \Big[ 
 w_t \left\| \vec{t}_{\mathrm{sim}}^n - \vec{t}_{\mathrm{exp}}^n \right\|^2
+ w_f \left\| \vec{F}_{\mathrm{sim}}^n - \vec{F}_{\mathrm{exp}}^n \right\|^2 \nonumber \\
& \ \ + w_d \left\| T_{\mathrm{sim}}^n - T_{\mathrm{exp}}^n \right\|^2
+ w_r \left\| \vec{R}^n(\vec{q}) \right\|^2  \Big],
\end{align}
where $ \vec{t}_{\mathrm{sim}}^n $, $ \vec{F}_{\mathrm{sim}}^n $, and $ T_{\mathrm{sim}}^n $ denote the simulated footfall timing vector, tugline force trajectory, and stride duration for stride $ n $, respectively. Experimental data are denoted by the corresponding $ \mathrm{exp} $ subscripts. The residual vector $ \vec{R}^n(\vec{q}) $ captures physical consistency constraints, including both (i) holonomic constraints enforcing correct leg lengths at touchdown and liftoff events, and (ii) continuity constraints for state variables across stride boundaries. The weights $ w_r, w_t, w_f, w_d $ are manually tuned to balance the influence of each term.

\textbf{Single-Stride Optimization:} When $ N = 1 $, the optimization produces a periodic gait by enforcing cycle closure constraints:
\[
\vec{q}(t_0) = \vec{q}(t_{\mathrm{end}}), \quad \dot{\vec{q}}(t_0) = \dot{\vec{q}}(t_{\mathrm{end}}),
\]
ensuring that the simulated trajectory forms a dynamically consistent limit cycle matching experimental stance patterns and tugline force profiles.

\textbf{Multi-Stride Optimization:} For gait transitions and sequence replication, the trajectory is segmented into $ N $ strides indexed by $ n = 1{:}N $, with each stride spanning $ t_0^n $ to $ t_{\mathrm{end}}^n $. Continuity is enforced between successive strides:
\[
\vec{q}(t_0^{n+1}) = \vec{q}(t_{\mathrm{end}}^n), \quad \dot{\vec{q}}(t_0^{n+1}) = \dot{\vec{q}}(t_{\mathrm{end}}^n).
\]
To replicate experimentally observed stride-to-stride gait changes, the swing stiffness parameters $ k_{\mathrm{swing},i}^n $ for $ i \in \{\small\text{LH}, \small\text{LF}, \small\text{RF}, \small\text{RH}\} $ are treated as stride-specific optimization variables and allowed to vary across strides.
Initial guesses are manually constructed based on approximate periodic solutions and refined using a direct collocation method combined with nonlinear programming. This procedure yields dynamically feasible trajectories that closely reproduce experimental stride patterns and tugline force behaviors for both steady-state and transitioning gaits.

\section{Results}
\label{sec: results}

In this section, we present the analysis of general patterns observed in galloping gaits used by load dogs during load pulling. We observed that load dogs frequently and randomly transition among four different types of galloping gaits. To reveal the underlying mechanism behind these frequent and random gait transitions, we applied our quad-load model optimization approach to achieve two primary objectives. First, we aimed to replicate the quad-load motion accurately for a single stride, demonstrating our model's capability to reproduce realistic locomotion dynamics. Second, we conducted simulations replicating the quad-load motion across different galloping gait types. In this second part, selected model parameters were allowed to vary freely between strides, enabling us to uncover the mechanisms driving gait transitions by analyzing parameter variations.

\subsection{Observed Distribution of Gait Patterns}
\label{subsec: statistics}
\begin{figure*}
    \centering
    \includegraphics[width=0.99\textwidth]{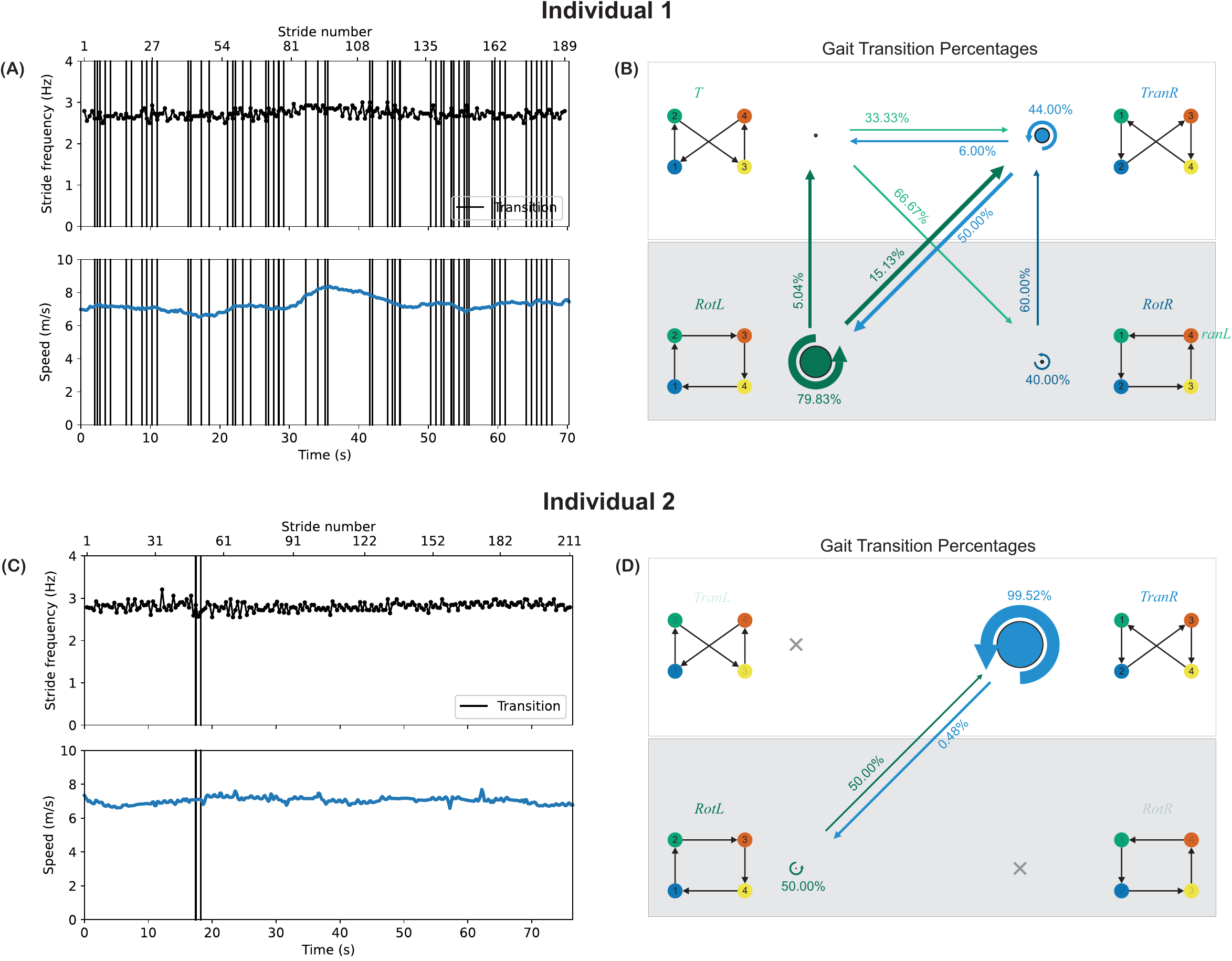} 
    \caption{
    Stride frequency, forward speed, and gait transitions for Individual 1 (A,B) and Individual 2 (C,D). 
    (A,C) Time series of stride frequency (top) and forward speed (bottom). Stride frequency was computed as the inverse of the interval between consecutive touchdowns of the trailing hind limb (upper x-axis indicates stride number). Forward speed was estimated from towing-vehicle speed (5~s centered moving average). Vertical grey lines mark identified gait transitions. Gait labels are summarized in Table~\ref{tab:footfall_r2_summary}.
    (B,D) Gait usage and transition network.    
    Each node represents a distinct gait type, with the dot size proportional to its overall usage frequency. Directed arrows indicate transitions between gaits, with arrow thickness representing transition frequency. For each gait node, a corresponding dot plot is shown to illustrate the footfall sequence, using the same color (green for left-leading, blue for right-leading) and background (white for transverse, grey for rotary) convention as defined in Fig.~\ref{fig: galloping def 2}. This integrated view links gait prevalence, transitions, and sequence structure in a unified representation.
    Note that the transition network is based on finite observations (Individual~1: 189 strides with 60 transitions; Individual~2: 211 strides with 2 transitions), and several specific gait types or transition classes are supported by few or zero samples; thus, low-frequency arrows (or missing links) should be interpreted as sample-limited rather than definitive.
    }
    \label{fig:gait_statistics}
\end{figure*}

Our analysis of high-speed video recordings revealed that sled dogs frequently switch between different galloping gaits during load-pulling.
Notably, these gait transitions occurred across relatively narrow speed and stride-frequency ranges. GPS traces from the towed vehicle indicated that both trials were performed at similar forward speeds (Fig.~\ref{fig:gait_statistics}). Individual~1 ran at a mean speed of 7.28~m/s (range 6.52--8.38~m/s; CV $= 5.6$\%), while Individual~2 ran at a mean speed of 7.03~m/s (range 6.57--7.69~m/s; CV $= 2.7$\%). Within each trial, speed fluctuated modestly around the mean. Stride frequency likewise remained relatively constant (Individual~1: 2.72~Hz; Individual~2: 2.81~Hz). Transitions occurred even when speed was relatively constant, giving no indication that transitions were restricted to bouts of acceleration or deceleration.
To assess stride regularity, we quantified stance and swing durations for each limb. For Individual~1, stance durations were tightly distributed (pooled across limbs: mean $=0.0844$~s, SD $=0.0074$~s, CV $=8.8\%$), whereas swing durations were longer and more variable (mean $=0.2836$~s, SD $=0.0263$~s, CV $=9.3\%$), consistent with gait transitions being implemented via interlimb phase adjustments during swing (i.e., lead-limb swaps in flight) rather than through delays or irregularities in stance timing. Individual~2 exhibited a similar stance distribution (mean $=0.0877$~s, SD $=0.0079$~s, CV $=9.0\%$), while swing durations were slightly shorter and less variable (mean $=0.2689$~s, SD $=0.0124$~s, CV $=4.6\%$), consistent with the lower transition rate in this trial.

Fig.~\ref{fig:gait_statistics} also summarizes the observed distribution of gallops and the frequency of transitions between them for two individual dogs. The arrangement of gait types in the figure follows the same convention as Fig.~\ref{fig: galloping def 2}, with transverse gaits (\textit{TranL}, \textit{TranR}) placed above and rotary gaits (\textit{RotL}, \textit{RotR}) below, and left- and right-leading gaits positioned to the left and right, respectively.
The size of each dot is proportional to the number of strides recorded for that gait, while the thickness of the connecting arrows reflects the frequency of transitions between gaits. 

Data for Individual 1, compiled from 189 recorded strides, reveals the use of all four galloping gait patterns. Among these, \textit{RotL} was the most frequently used, occurring in 120 out of 189 strides. This was followed by \textit{TranR}, which appeared in 50 strides. In contrast, \textit{TranL} and \textit{RotR} were used only occasionally.
Gait transitions were unevenly distributed. For \textit{RotL}, 79.83\% of consecutive strides remained in the same gait. Transitions to \textit{TranR} occurred in 15.13\% of cases, and to \textit{TranL} in 5.04\%. 
For strides beginning in \textit{TranR}, 44.00\% remained in \textit{TranR}, 50.00\% transitioned to \textit{RotL}, and only 6.00\% transitioned to \textit{TranL}. No transitions to \textit{RotR} were observed from \textit{TranR} either.
These statistics suggest a strong preference for \textit{RotL} and \textit{TranR}, while \textit{TranL} and \textit{RotR} appear to be infrequent or short-lived. The most common transitions occurred between \textit{RotL} and \textit{TranR} (\textit{RotL} $\leftrightarrow$ \textit{TranR}), which require changing the touchdown order of only the hindlimbs. In contrast, transitions between \textit{RotL} and \textit{RotR} were never observed.

Data for Individual 2, compiled from 211 recorded strides, reveals a highly consistent gait preference. Nearly all strides were executed in the \textit{TranR} gait, with only a small fraction performed in \textit{RotL}. Specifically, \textit{TranR} accounted for 209 out of 211 strides, while \textit{RotL} was observed in just two strides. The other two gaits, \textit{TranL} and \textit{RotR}, were not used at all in the recorded dataset.
No transitions were observed among three of the four gait types, as \textit{TranR} and \textit{RotL} were the only patterns present. 
For strides beginning in \textit{TranR}, 99.52\% remained in the same gait, while only 0.48\%, corresponding to a single stride, transitioned to \textit{RotL}. 
For \textit{RotL}, only two strides were observed, corresponding to one self-transition  and one transition to \textit{TranR}.
This consistency indicates a strong and stable preference for the \textit{TranR} gait in Individual 2, with other patterns either avoided or unsustainable. The lack of bidirectional transitions suggests that, in contrast to Individual 1, this dog maintained a highly stable coordination strategy with minimal gait variability across the recorded session.

Although Fig.~\ref{fig:gait_statistics} summarizes both gait occupancy and the observed transition graph, the number of samples supporting several specific transition types is small, and the corresponding transition-rate estimates should be interpreted cautiously. For Individual~1, the overall transition count was 60 transitions across 189 recorded strides, and even the most frequently observed directed transitions were supported by a limited number of events (e.g., \textit{TranR}\,$\rightarrow$\,\textit{RotL}: $n=25$; \textit{RotL}\,$\rightarrow$\,\textit{TranR}: $n=18$), while many other transition categories occurred only a handful of times or were not observed. For Individual~2, the gait distribution was even more concentrated: \textit{TranR} accounted for 209/211 strides and transitions occurred in only 2 strides, with \textit{RotL} observed only twice and no occurrences of \textit{TranL} or \textit{RotR}. Consequently, any inferences about fine-grained preferences among rare gait types, or about the relative likelihood of specific low-frequency transitions, are limited by the available sample sizes. For this reason, the remainder of this work emphasizes modeling the coupled load-pulling dynamics of the dog--sled system and evaluating whether the observed transition phenomena can be replicated, rather than drawing strong statistical conclusions from sparsely sampled transition classes.

\subsection{Single Stride Replication}

In this subsection, we applied the optimization algorithm described in Section~\ref{sec:opti} to replicate the four galloping gaits observed in our dataset. We compared footfall timings and tugline forces for both Individual~1 and Individual~2 across the gaits \textit{RotL}, \textit{RotR}, \textit{TranL}, and \textit{TranR}.

\begin{figure*}
    \centering
    \includegraphics[width=0.85\textwidth]{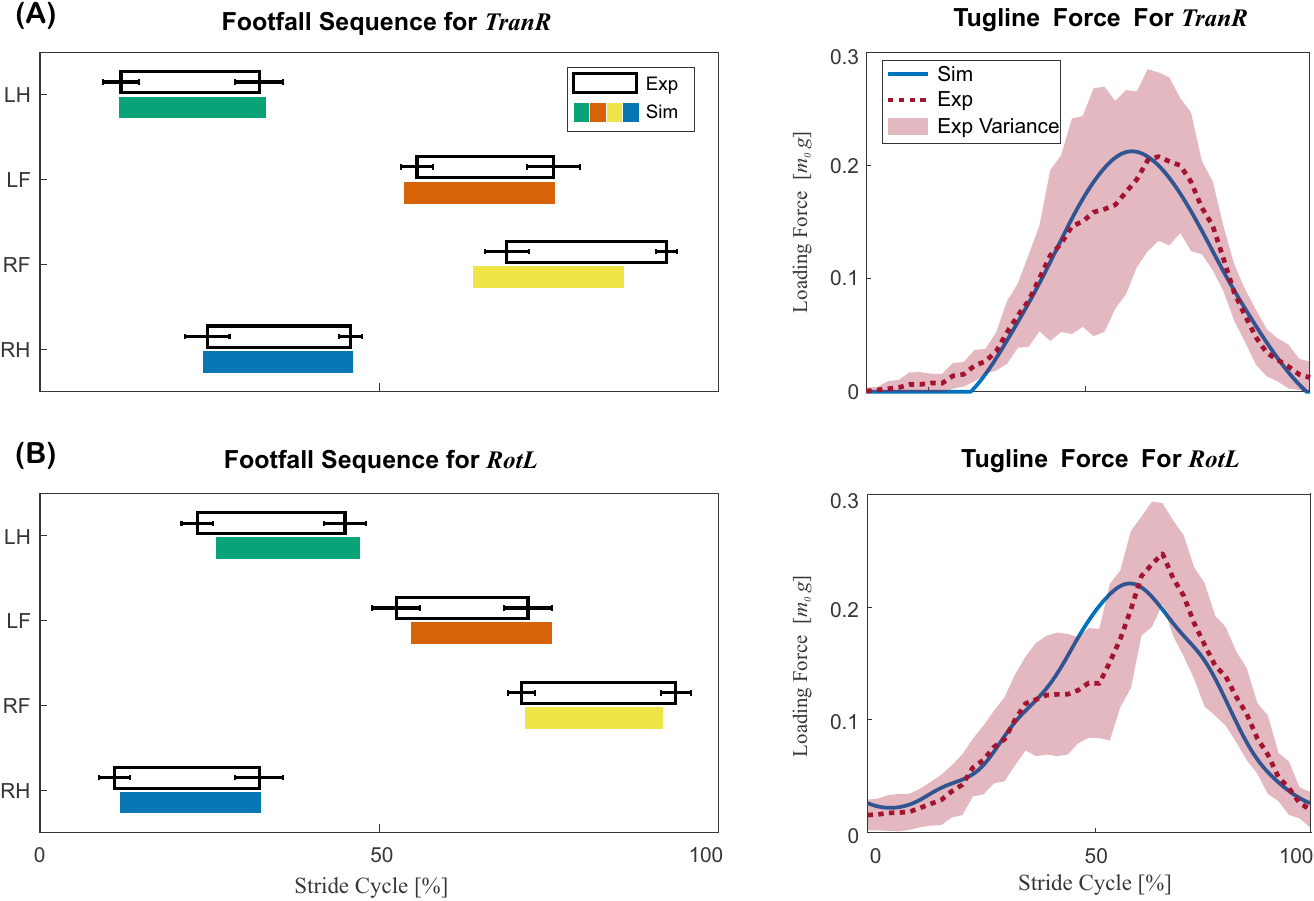} 
    \caption{Single-stride solutions for Individual 1, comparing two gaits: (A) \textit{TranR} and (B) \textit{RotL}. Left subplots show experimental and simulated footfall sequences; right subplots illustrate experimental and simulated tugline forces along the tugline. Experimental means (dashed curves) and variances (shaded regions and error bars) are computed from 50 (\textit{TranR}) and 120 (\textit{RotL}) samples. }
    \label{fig:single_stride}
\end{figure*}

For Individual~1, the two most frequently used gaits, \textit{RotL} and \textit{TranR}, are shown in Fig.~\ref{fig:single_stride} for detailed comparison between empirical observations and model simulation. Mean touchdown and liftoff timings were extracted from the recorded stride sequences. The averaged stance durations for each leg are shown as white bars in the left panels of Fig.~\ref{fig:single_stride}, with error bars indicating variability. Simulated stance durations are overlaid as colored bars for direct comparison. In the right panels, tugline forces are shown as solid blue curves for the simulation and as dashed red curves with shaded bands indicating the measured mean and variance.

\begin{table}
\setlength\tabcolsep{4pt} 
\centering
\caption{Summary of $R^2$ evaluations for footfall timing ($R^2_t$), tugline force ($R^2_F$), and stride duration ($R^2_D$) across different gait types and individuals. The number of strides used to compute each experimental average is listed in the rightmost column. Data for \textit{RotL} and \textit{TranR} for Individual 1 correspond to those shown in Fig.~\ref{fig:single_stride}.}
\label{tab:footfall_r2_summary}
\begin{tabular}{>{\columncolor{white}}l|c|c c c|c}
\hline
Dog Name   &   Gait Type   &   $R^2_T$   &   $R^2_F$   &   $R^2_D$   &  \# Strides \\
\hline
Individual 1   &   \textit{TranL}   &   0.99   &   0.95   &   0.89   & 9 \\
\rowcolor{gray!15}
Individual 1   &   \textit{TranR}   &   0.99   &   0.96   &   0.99   & 50 \\
\rowcolor{gray!15}
Individual 1   &   \textit{RotL}   &   0.99   &   0.94   &   0.93   & 120 \\
Individual 1   &   \textit{RotR}   &   0.99   &   0.96   &   0.99   & 10 \\
Individual 2   &   \textit{TranR}   &   0.98   &   0.96   &   0.84   & 209 \\
Individual 2   &   \textit{RotL}   &   0.98   &   0.92  &   0.87   & 2 \\
\hline
\end{tabular}
\end{table}

For the \textit{RotL} gait of Individual~1 (Fig.~\ref{fig:single_stride}A), the simulated stance onset and offset closely match the experimental profiles across all limbs, indicating precise replication of footfall timing. Quantitatively, the mean stance duration error remains below 3\% of the stride cycle for each leg. The tugline force profile also demonstrates strong agreement: the simulated curve captures both the timing and amplitude of the experimental force peaks, with a peak value reaching approximately $0.27\,mg$, consistent with the observed average. Although minor discrepancies occur during the force rising phase, they remain within the bounds of experimental variability. This high level of fidelity is reflected in the quantitative metrics summarized in Table~\ref{tab:footfall_r2_summary}, where the \textit{RotL} simulation achieves $R^2_T = 0.99$ for footfall timing, $R^2_F = 0.94$ for tugline force, and $R^2_D = 0.93$ for stride duration.

The model performs similarly well for the \textit{TranR} gait of Individual~1 (Fig.~\ref{fig:single_stride}B), further demonstrating its robustness across gait types. Stance durations are accurately reproduced, and notable asymmetries in forelimb stance timing—evident in the experimental data—are qualitatively reflected in the simulation. Tugline force agreement remains strong, with an overall $R^2_F = 0.96$, and the simulated peak force amplitude closely aligns with the experimental mean. In terms of timing accuracy, the footfall sequence achieves an excellent match with $R^2_T = 0.99$. Furthermore, the stride duration agreement is nearly perfect, as indicated by a coefficient of determination of $R^2_D = 0.99$.

Due to space constraints, the optimization results for the remaining gaits and for Individual 2 are not visualized here but are included in the publicly available source code and data repository (see the Data and materials availability section).
Table~\ref{tab:footfall_r2_summary} summarizes the $R^2$ values across all gait and individual combinations. Footfall timing predictions are highly consistent, with all combinations achieving $R^2_t = 0.99$, indicating precise alignment of simulated and experimental touchdown and liftoff events. Tugline force predictions also show strong agreement, with $R^2_F$ values ranging from 0.95 to 0.96 across gaits. Stride duration tracking is most accurate for the frequently used \textit{RotL} and \textit{TranR} gaits, where $R^2_D$ exceeds 0.99, and is slightly lower for \textit{TranL} and \textit{RotR}, which exhibit more variability.
Although \textit{TranL} and \textit{RotR} are less frequently used gaits, the SLIP model still maintains reasonable fidelity, with timing and force $R^2$ values above 0.93 in all cases. Slightly reduced agreement in \textit{TranL}, especially for Individual 1, may reflect increased variability in experimental execution or decreased model robustness for rarely used gaits. Nevertheless, the model performs reliably across all conditions tested.

\begin{figure*}[htb]
    \centering
    \includegraphics[width=0.95\textwidth]{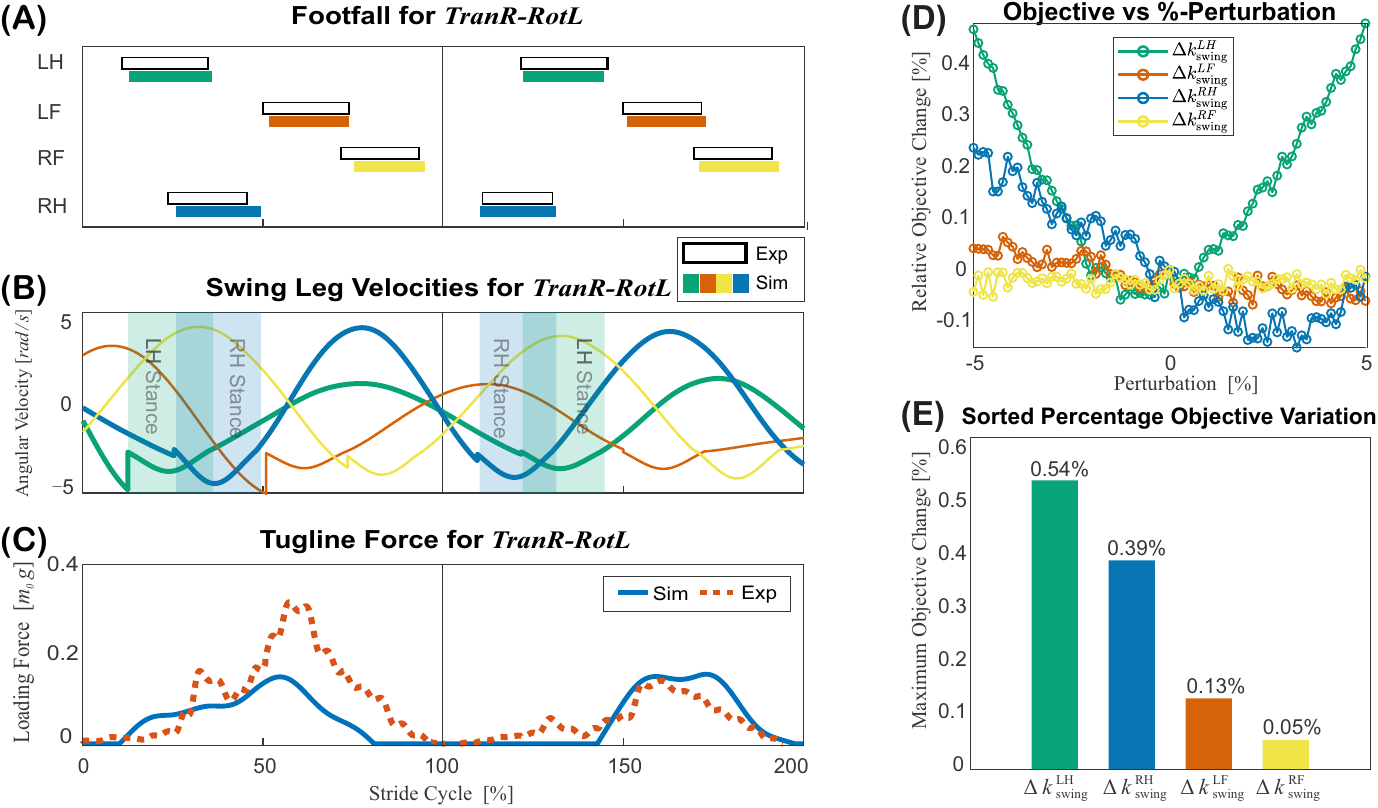} 
    \caption{Replicated gait transition for Individual~1 from transverse right (\textit{TranR}) to rotary left (\textit{RotL}) gallop.  
    \textbf{(A)} Simulated footfall sequence. 
    \textbf{(B)} Swing leg angular velocities. 
    \textbf{(C)} Tugline force trajectory. 
    \textbf{(D)} Absolute cost variation under $\pm5\%$ perturbations in $\Delta k_{\text{swing}, i}$. 
    \textbf{(E)} Relative percentage cost variation. 
    These results highlight the dominant role of hindlimb stiffness modulation in driving the transition.}
    \label{fig:gait_transition 1}
\end{figure*}

\subsection{Gait Transition Replication}
\label{subsec:Repro_Transition}

The previous subsection demonstrated that our SLIP-based model accurately reproduces footfall timings and tugline forces across all four galloping gaits, with especially high fidelity for the most frequently used patterns, \textit{RotL} and \textit{TranR}. Building on these single-stride results, we now examine whether the model can also capture stride-to-stride gait transitions, which occur frequently in the experimental recordings. As detailed in Section~\ref{subsec: statistics}, the \textit{TranR} gait was the dominant pattern for both Individual~1 and Individual~2, with \textit{TranR}$\rightarrow$\textit{RotL} accounting for the majority of observed transitions in Individual~1. Transitions involving \textit{RotR} and \textit{TranR} were also recorded, though less commonly. To evaluate whether these experimentally observed transitions can be replicated through targeted modulation of limb-specific swing stiffness, we applied our trajectory optimization framework to two-stride sequences extracted from continuous locomotion data, focusing on the \textit{TranR}$\leftrightarrow$\textit{RotL} and \textit{RotR}$\leftrightarrow$\textit{TranR} transitions.

\begin{table}
\setlength\tabcolsep{4pt}
\centering
\caption{Swing stiffness changes ($ \Delta k_{\mathrm{swing},\ i} := k_{\mathrm{swing},\ i}^{n+1} - k_{\mathrm{swing},\ i}^n $  $ i \in \{\small\text{LH}, \small\text{LF}, \small\text{RF}, \small\text{RH}\} $) across transitions. Highlighted cells indicate the leg pair responsible for the transition.}
\label{tab:transitions_delta}
\footnotesize
    \begin{tabular}{l c c r r r r}
        \toprule
        \textbf{Transition} & \textbf{Switched} & $R^2_{\text{w}}$ & LH & RH & LF & RF \\
        \midrule
        \textit{TranR}$\to$\textit{RotL} & Hind Pair & 0.85
        & \cellcolor[gray]{0.85}$-23.8$ & \cellcolor[gray]{0.85}$+23.1$ 
        & $-1.38$ & $0.57$ \\
        \textit{RotL}$\to$\textit{TranR} & Hind Pair & 0.88
        & \cellcolor[gray]{0.85}$+12.5$ & \cellcolor[gray]{0.85}$-17.7$ 
        & $-4.10$ & $0.72$ \\
        \textit{RotL}$\to$\textit{TranL} & Fore Pair & 0.96
        & $-1.80$ & $-0.50$ 
        & \cellcolor[gray]{0.70}$-9.5$ & \cellcolor[gray]{0.70}$15.2$ \\
        \bottomrule
    \end{tabular}
\end{table}

\subsubsection*{\textit{TranR}~$\leftrightarrow$~\textit{RotL}
 Gait Transition}
\label{subsubsec:TR-RL}

As shown in Fig.~\ref{fig:gait_transition 1}(A), the simulation accurately reproduces the key mechanical features of the \textit{TranR}~$\rightarrow$~\textit{RotL} transition, achieving a weighted coefficient of determination of $ R^2_{\text{w}} = 0.85 $. Both the footfall sequence and the tugline force trajectory closely match experimental observations across the two-stride transition. The middle panel illustrates swing leg angular velocities, which exhibit near-sinusoidal trajectories during flight—consistent with passive leg oscillations. Since the oscillation period inversely reflects swing stiffness, shorter cycles indicate higher values of $ k_{\text{swing}} $.

A clear reversal in hindlimb asymmetry is observed across the transition. In the first stride, the left hind (LH, green) shows a shorter period than the right hind (RH, blue), suggesting higher stiffness. In the second stride, this pattern reverses, with RH now oscillating faster than LH. The forelimb trajectories (LF: red, RF: yellow), by contrast, remain symmetric between strides, indicating stable stiffness. These observations suggest that the transition is primarily driven by differential modulation of hindlimb swing stiffness.

Quantitative analysis supports this interpretation. Defining the stride-to-stride stiffness change as $ \Delta k_{\mathrm{swing},\ i} := k_{\mathrm{swing},\ i}^{n+1} - k_{\mathrm{swing},\ i}^n $, we observe substantial and opposing variations in the hindlimbs: $ \Delta k_{\mathrm{swing},\ \scriptstyle\text{LH}} = -23.8\,mgl_0 $ and $ \Delta k_{\mathrm{swing},\ \scriptstyle\text{RH}} = +23.1\,mgl_0 $. In contrast, the forelimbs exhibit only minor adjustments: $ \Delta k_{\mathrm{swing},\ \scriptstyle\text{LF}} = -1.38\,mgl_0 $ and $ \Delta k_{\mathrm{swing},\ \scriptstyle\text{RF}} = +0.57\,mgl_0 $. These values align with the observed swing leg velocity asymmetries and are summarized in Table~\ref{tab:transitions_delta}, where the dominant hindlimb contributions are highlighted.

To evaluate whether the identified stiffness pattern is both necessary and sufficient for achieving the transition, we performed a sensitivity analysis by perturbing each $ \Delta k_{\text{swing}} $ by $\pm5\%$ independently and observing changes in the objective function. 
As shown in Fig.~\ref{fig:gait_transition 1}(D, top), the hindlimb curves (LH and RH) are consistent, in aggregate, with a local minimum at zero perturbation, indicating local optimality with respect to that parameter direction.
In contrast, the forelimb curves (LF and RF) exhibit much smaller variations. The maximum cost increase was 0.54\% for LH and 0.39\% for RH, compared to only 0.13\% and 0.05\% for LF and RF, respectively. These results confirm that the transition is highly sensitive to hindlimb stiffness modulation, while the forelimbs remain largely uninvolved.

To further test the sufficiency of this mechanism, we simulated the reverse transition (\textit{RotL}~$\rightarrow$~\textit{TranR}). The resulting stiffness changes were $ +12.5\,mgl_0 $ for LH and $ -17.7\,mgl_0 $ for RH, again showing large, opposing adjustments in the hindlimbs. The forelimb changes remained small: $ -4.1\,mgl_0 $ for LF and $ +0.72\,mgl_0 $ for RF. These results mirror the asymmetry observed in the forward transition and reinforce the conclusion that abrupt hindlimb stiffness modulation is sufficient to switch footfall sequences. The full set of values is provided in Table~\ref{tab:transitions_delta}, using the same highlighting scheme.

Together, these results show that the \textit{TranR}~$\leftrightarrow$~\textit{RotL} transition can be achieved through hindlimb stiffness modulation alone, without requiring adjustments in the forelimbs. This supports the broader hypothesis that discrete gait transitions in quadrupeds can be driven by targeted, stride-scale mechanical adjustments in a small subset of limbs.

\subsubsection*{\textit{RotL}~$\leftrightarrow$~\textit{TranL} Gait Transition}
\label{subsubsec:RL-TranL}

\begin{figure*}[!t]
    \centering
    \includegraphics[width=0.95\textwidth]{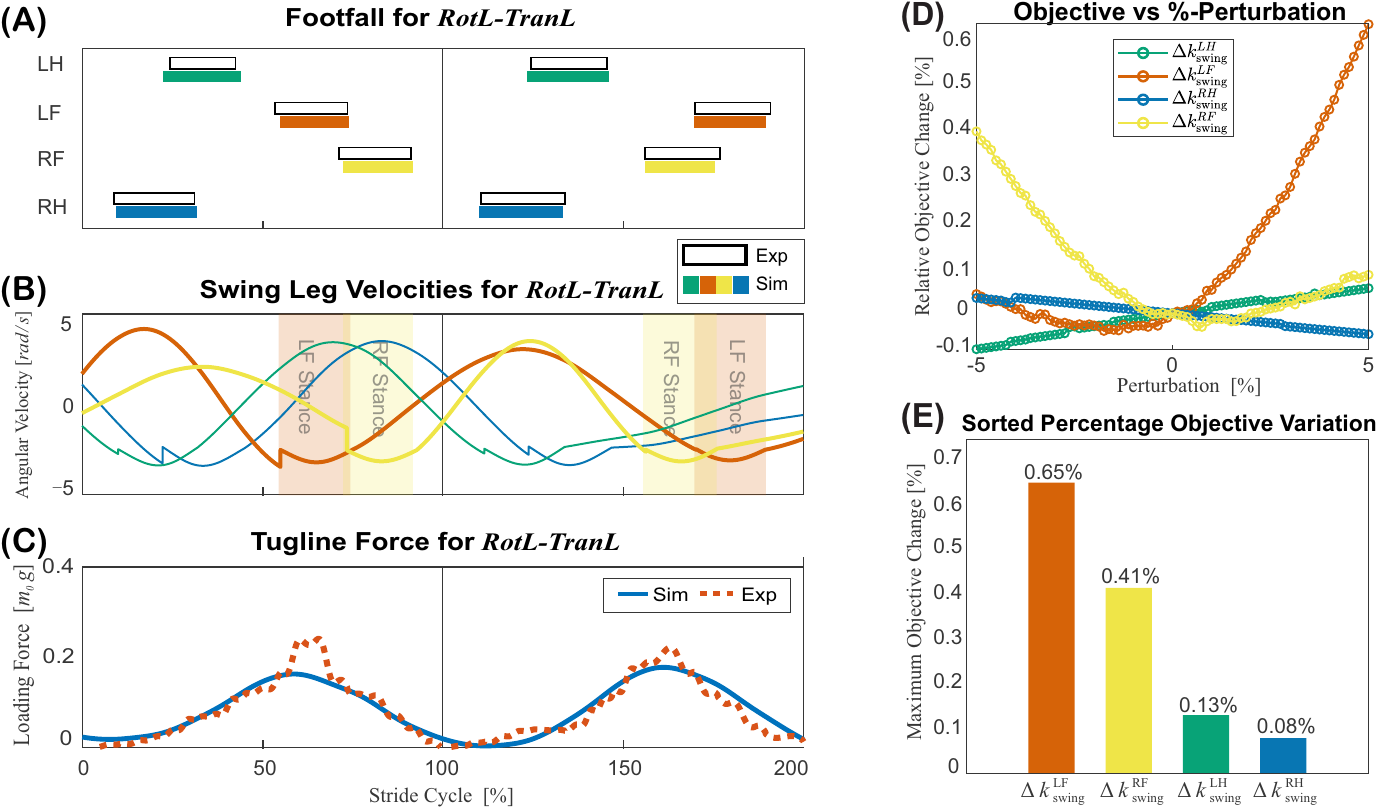} 
    \caption{Replicated gait transition for Individual~1 from rotary left (\textit{RotL}) to transverse left (\textit{TranL}) gallop.  
    \textbf{(A)} Simulated footfall sequence. 
    \textbf{(B)} Swing leg angular velocities. 
    \textbf{(C)} Tugline force trajectory. 
    \textbf{(D)} Absolute cost variation under $\pm5\%$ perturbations in $\Delta k_{\text{swing}, i}$. 
    \textbf{(E)} Relative percentage cost variation. 
    These results highlight the dominant role of forelimb stiffness modulation in enabling the transition. }
    \label{fig:gait_transition 2}
\end{figure*}

Figure~\ref{fig:gait_transition 2}(A) illustrates the simulated dynamics for the \textit{RotL}~$\rightarrow$~\textit{TranL} transition. The model achieves excellent agreement with experimental footfall and tugline force data, yielding a weighted fit score of $ R^2_{\text{w}} = 0.96 $. In the middle panel, the left fore (LF, red) initially oscillates faster than the right fore (RF, yrllow), and this asymmetry increases in the second stride—RF becomes even faster, while LF slows. In contrast, the hindlimb angular velocities remain nearly identical across both strides, suggesting that the transition is primarily driven by forelimb stiffness modulation.

This interpretation is supported by the corresponding changes in swing stiffness (Table~\ref{tab:transitions_delta}). The forelimbs exhibit strong, opposing changes: $ \Delta k_{\mathrm{swing},\ \scriptstyle\text{LF}} = -9.5\,mgl_0 $ and $ \Delta k_{\mathrm{swing},\ \scriptstyle\text{RF}} = 15.2\,mgl_0 $. The hindlimb changes are minimal: $ \Delta k_{\mathrm{swing},\ \scriptstyle\text{LH}} = -1.8\,mgl_0 $ and $ \Delta k_{\mathrm{swing},\ \scriptstyle\text{RH}} = -0.5\,mgl_0 $. This pattern, highlighted in the table with shaded forelimb cells, contrasts sharply with the hindlimb-driven transition observed in the \textit{TranR}~$\leftrightarrow$~\textit{RotL} case.
Sensitivity analysis (Fig.~\ref{fig:gait_transition 2}(B)) confirms the dominant role of the forelimbs. Perturbations of $ \Delta k_{\mathrm{swing},\ i} $ in LF and RF led to cost variations of 0.65\% and 0.41\%, respectively, compared to just 0.13\% and 0.08\% for LH and RH. These findings demonstrate that precise modulation of forelimb swing stiffness is sufficient to reproduce the observed footfall sequence transition.

Since the reverse transition (\textit{TranL}~$\rightarrow$~\textit{RotL}) was not observed in our dataset, we limited our analysis to the \textit{RotL}~$\rightarrow$~\textit{TranL} case. Nonetheless, the consistency of the stiffness modulation and the high reconstruction fidelity reinforce our broader hypothesis that discrete gait transitions can be achieved through stride-scale adjustments to a subset of limb parameters, in this case via targeted forelimb stiffness adaptation.

\section{Discussion and Conclusion}

\subsubsection*{Gait transitions in high speed galloping}

Gait transitions in quadrupeds are typically associated with changes in locomotor speed. High-speed galloping is often viewed as exhibiting consistent, stereotyped stride patterns~\cite{Hildebrand1977, Hilderbrand1989Quadrupedal, Hoyt1981gaitenergetic}. In contrast, our analysis of sprint sled dogs reveals frequent, stride-to-stride transitions among distinct galloping gaits that occur without significant changes in speed, stride duration, or terrain. 
This interpretation is consistent with symmetry-based theories of locomotion, in which morphology and symmetry structure shape an effective potential landscape with multiple local minima corresponding to distinct gaits \cite{Wilshin2017,Golubitsky1999}. Although the present quad-load SLIP model is not explicitly formulated as a gradient system, our results are compatible with this perspective. For fixed speed and load, we identify distinct periodic solutions corresponding to different galloping gaits, with sensitivity analyses (Figs.\ref{fig:gait_transition 1}D and \ref{fig:gait_transition 2}D) revealing locally well-defined optima associated with each solution. These optima can be interpreted as coexisting attractors in an effective locomotor landscape shaped by swing-leg stiffness asymmetries. Within this framework, gait transitions may arise from perturbation-driven transitions between neighboring attractors rather than from explicit changes in speed or global control strategy.
This behavior was especially prominent in Individual~1, who alternated between gait types under nearly identical locomotor conditions. This observation brings into question the role of multistability in quadrupedal load pulling. Prior studies have suggested that unloaded quadrupeds may switch lead limbs within an asymmetrical gallop to help balance muscular workload and mitigate fatigue~\cite{walter2007ground}. In sustained load-pulling tasks, the ability to switch between mechanically distinct gaits would potentially provide sled dogs a functional advantage for maintaining long-duration performance. 
Fatigue was not measured here, but both individuals were recorded during comparable training sessions, so we do not consider it an obvious explanation for why Individual~1 switched frequently while Individual~2 remained largely in a single gait.

It has also been suggested that sufficiently large perturbations might induce transitions between distinct gaits, shifting coordination toward patterns consistent with increased stability~\cite{wilshin2017longitudinal}. Indeed, it has been shown that controlled substrate perturbations can transiently bias gait selection in mice~\cite{vahedipour2018uncovering}.
Along these lines, it has been demonstrated that load-dependent phase-response rules that bias limbs toward stance can generate and emergently transition between quadrupedal gait patterns~\cite{owaki2017quadruped, zhang2024learning}. Such phase-response mechanisms represent a plausible route by which tugline perturbations could bias coordination and promote transitions among nearby stable galloping solutions in sled dogs. However, our data constrain how such a controller can be expressed: stance timing remained tightly distributed across limbs in both individuals, with transitions expressed via interlimb phase adjustments during swing/flight rather than through stance-time modulation. Thus, mechanisms requiring stance-timing disruptions---even loss of traction---do not align with the observed stride dynamics.
Irrespective of the specific mechanism, the ability to switch between mechanically distinct galloping solutions under steady-speed load pulling may still provide functional flexibility during sustained work.
However, the consistent gait observed in Individual~2 suggests that such transitions, if they exist, are not strictly necessary. 
Notably, both individuals primarily exhibited \textit{Tran}$\leftrightarrow$\textit{Rot} transitions, whereas within-family lead switches (e.g., \textit{RotL}$\leftrightarrow$\textit{RotR}) were more rare or absent in our dataset. Within-family switches require a simultaneous reordering of both the forelimb and hindlimb touchdown sequences within a single stride, rather than a single limb pair. This more complex reconfiguration may impose greater coordination demands at high speeds.

\subsubsection*{A minimal model for reproducing stride-to-stride transitions}

To understand the mechanisms underlying these transitions, we developed a reduced-order SLIP model with a load-coupled body. Using footfall timing and tugline force data as references, we applied trajectory optimization to replicate all observed galloping gaits and their stride-to-stride transitions. Despite its simplicity, the model successfully reproduced both steady-state gaits and transition dynamics using a small number of adjustable stride-level parameters.

Among the parameters examined, swing leg stiffness emerged as a dominant control variable for enabling gait transitions. Biomechanically, this parameter governs how rapidly a leg swings through the air: higher stiffness corresponds to faster oscillation and earlier touchdown. Our results demonstrate that targeted modulation of swing stiffness in just one limb pair was sufficient to replicate the experimentally observed transitions. Specifically, modulating hindlimb stiffness alone reproduced the transition between \textit{TranR} and \textit{RotL}, while adjusting forelimb stiffness enabled the transition from \textit{RotL} to \textit{TranL}. 
These adjustments effectively alter the sequencing of leg touchdown events and thus the overall footfall pattern. Sensitivity analyses further confirmed that the cost function was highly responsive to perturbations in the switched limb pair and relatively insensitive to changes in the other limbs. This suggests that sled dogs may exploit stride-wise adjustments in swing timing—achieved through limb-specific stiffness control—as a localized, efficient strategy to switch gaits without global reconfiguration or speed change.

These findings extend previous SLIP-based approaches by showing that transitions between distinct galloping gaits can be replicated without speed variation or terrain change. Instead, stride-scale adjustments to mechanical parameters in a small set of limbs appear sufficient to drive discrete footfall sequence changes, even under external loading.

\subsubsection*{Limitations and future directions}

Although our model replicates footfall timings and tugline dynamics with high accuracy, the underlying causes for the observed transitions remain unresolved. One hypothesis is that transitions are influenced by fluctuations in tugline force. However, across 120 \textit{RotL} strides from Individual~1 and 179 \textit{TranR} strides from Individual~2, most strides—79\% and 99\% respectively—did not involve a transition. 
Importantly, the present dataset is not adequate to quantify the driving mechanisms of gait switching. The number of observed transitions is too small to estimate load- or speed-dependent transition probabilities with confidence. Additional data is necessary to test whether and how mechanical perturbations bias gait switching during high-speed load pulling.

In interpreting the transition fits and sensitivity results, several limitations should be kept in mind. First, our sensitivity analysis is centered on transition strides rather than mapping cost gradients around canonical TranR or RotL solutions, in part because the parameterization is effectively high-dimensional and over-actuated, so multiple parameter directions can yield acceptable fits and control-like gradients may not be unique.
Secondly, we used a model in which the primary dissipation arises from load friction and did not include explicit joint-level damping. Because damping can strongly influence task-level stability \cite{Heim2020} and may be especially consequential for transition strides, our conclusions regarding transition robustness and stability should be interpreted within this simplified structure.
Third, in Fig.~6, the cost is evaluated on a single two-stride transition instance, so we cannot strictly rule out instance-specific overfitting, even though similar qualitative stiffness-change structure appears across multiple realizations of the same transition type. 
More broadly, we do not claim that alternative parameters have been eliminated as contributors, and because the optimization is gradient-based we have not systematically characterized sensitivity to initial guesses or the structure of the optimization landscape.
Finally, stiffness modulation is one of the primary available levers to reorganize timing in our reduced-order model. While the results support a consistent model-based relationship between differential swing stiffness and footfall reordering, establishing biological causality would require predictive perturbations or independent validation.

Future work should use protocols that prescribe speed/acceleration profiles to quantify how transition probability depends on these factors. Although we restricted our analysis to straight, steady-running segments, speed was not experimentally controlled. Treadmill experiments, however, could enforce constant speeds, or---if the animal remains harnessed on the treadmill---could also apply controlled perturbations to quantify gait-switching responses.
Additionally, the short duration and moderate intensity of the bouts analyzed here preclude strong inference about fatigue-driven switching. Future experiments will require longer recordings and fatigue proxies (e.g., heart rate, metabolic estimates, or limb-loading asymmetries) to quantify whether lead-switching frequency increases with cumulative workload under load pulling.
In parallel, model extensions that incorporate limb-specific costs or fatigue-like dynamics could help interpret such experiments and generate testable predictions for when gait switching should emerge under load,
including whether within-family switches correspond to larger inferred costs than transverse--rotary switches.

Another limitation of this study lies in the inability to decouple the roles and contributions of individual dogs within the larger team. All experimental data presented here were collected from a ten-dog team hitched in a double-tandem (Alaska) configuration, where the distribution of tensions across the interconnected tether network is not directly measurable. This introduces some ambiguity in interpreting how load dynamics influence gait selection at the individual level. In practice, each dog's pulling contribution may vary depending on spatial position, terrain conditions, cornering behavior, fatigue, or even team coordination strategies. Whether such factors contribute to observed gait transitions remains an open question. Future studies could address this limitation by instrumenting each segment of line to measure complete hitch network tension in real time. Additionally, synchronized video and acceleration data from multiple team members could enable modeling of team-level coordination strategies, including how leading versus trailing dogs modulate their gait or force output in response to group dynamics. Decoupling these roles would allow for more precise attribution of observed gait transitions to local mechanical or environmental factors, and could clarify whether transitions are driven by individual control decisions or emergent team-level phenomena.

A primary barrier to addressing the limitations above is data scale. Rigorous transition statistics will require substantially more annotated strides across individuals and conditions. Although treadmill protocols can prescribe speed and perturbations in a controlled setting, field studies remain challenging both to deploy at scale and to annotate. Recent work, however, has demonstrated that stance and stride timing can be inferred from inertial measurement unit (IMU) data using machine learning approaches \cite{serra2020improving}, offering a valuable alternative when labeled video is limited or unavailable. 
To assess feasibility in our setting, we trained a bidirectional LSTM \cite{sak2014long} on Individual~1 using stance labels annotated from the synchronized 240~fps video and discretized onto the 120~Hz IMU time base. In a held-out test set (362 footfalls), the model recovered the correct footfall order and achieved an average intersection-over-union (IoU) of 0.869 for stance intervals. Typical discrepancies were on the order of one IMU timestep ($\sim$10~ms), near the effective temporal resolution of the sensor. Code and evaluation details are provided in our repository (Sec.~Data Availability). While promising, the current training set is still too limited for broader generalization. Nonetheless, IMU-only gait inference could substantially reduce reliance on continuous video and enable the larger datasets needed to quantify speed- and load-dependent transition probabilities.

\subsubsection*{Implications} 

Together, our stride-resolved measurements and SLIP modeling framework provide a new window into high-speed quadrupedal locomotion under sustained external loading. Our empirical results provide an additional example of discrete gait switching consistent with quadrupedal multistability, here observed at high speed during sustained load pulling. Although we do not identify a specific \emph{driver} of switching, sprint sled dogs alternate between transverse and rotary galloping patterns within a comparatively narrow speed range, suggesting that gait choice is not a deterministic function of speed alone.
It should be acknowledged that transverse--rotary switching corresponds to a relatively small reconfiguration of interlimb phasing compared with classical trot--gallop transitions Nevertheless, transitions among closely related galloping solutions are still consistent with multistable locomotor dynamics in a regime where limb timing is otherwise highly constrained, and may support robust locomotion under perturbation while redistributing muscular demand over time.
Our SLIP-based model reproduces observed switching using targeted modulation of a small subset of physically interpretable parameters, including stiffness asymmetries and swing-phase adjustments. This suggests that low-dimensional control knobs may be sufficient to organize gait switching in quadruped--load systems and providing a principled starting point for future, load-aware gait-switching controllers.

\section*{Animal Subjects and Ethics}
The data analyzed in this study were collected from two sprint-racing sled dogs (one male, one female; 1.5 and 2.5 years old, respectively) from a professional sprint-racing kennel. Both dogs were bred and trained for the sport of sprint racing and are representative of the Eurohound type. For the run analyzed here, dog positions within the team were chosen by their owner based on each dog's temperament and aptitude. All procedures were conducted in accordance with institutional and national guidelines for research involving animals. Protocols were reviewed and approved by the Institutional Animal Care and Use Committee at Georgia Institute of Technology (protocol BHAMLA-A100575U-10/06/2025). Written informed consent was obtained from the dogs’ owners prior to participation. Dogs were housed and cared for by their owners according to standard husbandry practices. All experimental sessions were integrated into the dogs’ normal training routine and were discontinued immediately if any signs of distress were observed.\\

\begin{acknowledgments}
We gratefully acknowledge the generous support of the sled dog mushers who volunteered their kennel and dogs for data collection. Their willingness to share their time, expertise, and home made this fieldwork possible.
\textbf{Funding:} S.B. acknowledges funding from NSF CAREER IOS-1941933 and Schmidt Sciences, LLC.
\textbf{Author contributions:}
Conceptualization: J.D., B.S. Experiments and related data analysis: B.S. 
Model and related analysis: J.D. 
Writing—original draft: J.D., B.S.
Writing—review and editing: J.D., B.S., S.B., Z.G.
Visualization: J.D., B.S.
Supervision, funding acquisition: S.B., Z.G.
\textbf{Competing interests:} The authors declare no competing interests.
\textbf{Data and materials availability:} All code to run simulations, generate figures, and reproduce results is publicly available at our online repository \texttt{https://github.com/DLARlab/2025\_Gait\_Transitions\\\_in\_Load\_Pulling\_Quadrupeds\_Insights.git}
\end{acknowledgments}

\bibliography{apssamp}

@article{kuo2005energetic,
  title={Energetic consequences of walking like an inverted pendulum: step-to-step transitions},
  author={Kuo, Arthur D and Donelan, J Maxwell and Ruina, Andy},
  journal={Exercise and sport sciences reviews},
  volume={33},
  number={2},
  pages={88--97},
  year={2005},
  publisher={LWW}
}

@article{shafiee2024viability,
  title={Viability leads to the emergence of gait transitions in learning agile quadrupedal locomotion on challenging terrains},
  author={Shafiee, Milad and Bellegarda, Guillaume and Ijspeert, Auke},
  journal={Nature Communications},
  volume={15},
  number={1},
  pages={3073},
  year={2024},
  publisher={Nature Publishing Group UK London}
}

@inproceedings{zhang2024learning,
  title={Learning emergent gaits with decentralized phase oscillators: on the role of observations, rewards, and feedback},
  author={Zhang, Jenny and Heim, Steve and Jeon, Se Hwan and Kim, Sangbae},
  booktitle={2024 IEEE International Conference on Robotics and Automation (ICRA)},
  pages={3426--3433},
  year={2024},
  organization={IEEE}
}

@article{owaki2017quadruped,
  title={A quadruped robot exhibiting spontaneous gait transitions from walking to trotting to galloping},
  author={Owaki, Dai and Ishiguro, Akio},
  journal={Scientific reports},
  volume={7},
  number={1},
  pages={277},
  year={2017},
  publisher={Nature Publishing Group UK London}
}

@article{vahedipour2018uncovering,
  title={Uncovering the structure of the mouse gait controller: Mice respond to substrate perturbations with adaptations in gait on a continuum between trot and bound},
  author={Vahedipour, A and Maghsoudi, O Haji and Wilshin, S and Shamble, P and Robertson, B and Spence, A},
  journal={Journal of biomechanics},
  volume={78},
  pages={77--86},
  year={2018},
  publisher={Elsevier}
}

@article{Wilshin2017,
  title = {Morphology and the gradient of a symmetric potential predict gait transitions of dogs},
  volume = {111},
  ISSN = {1432-0770},
  url = {http://dx.doi.org/10.1007/s00422-017-0721-2},
  DOI = {10.1007/s00422-017-0721-2},
  number = {3–4},
  journal = {Biological Cybernetics},
  publisher = {Springer Science and Business Media LLC},
  author = {Wilshin,  Simon and Haynes,  G. Clark and Porteous,  Jack and Koditschek,  Daniel and Revzen,  Shai and Spence,  Andrew J.},
  year = {2017},
  month = jun,
  pages = {269–277}
}

@article{wilshin2017longitudinal,
  title={Longitudinal quasi-static stability predicts changes in dog gait on rough terrain},
  author={Wilshin, Simon and Reeve, Michelle A and Haynes, G Clark and Revzen, Shai and Koditschek, Daniel E and Spence, Andrew J},
  journal={Journal of Experimental Biology},
  volume={220},
  number={10},
  pages={1864--1874},
  year={2017},
  publisher={The Company of Biologists Ltd}
}

@article{walter2007ground,
  title={Ground forces applied by galloping dogs},
  author={Walter, Rebecca M and Carrier, David R},
  journal={Journal of Experimental Biology},
  volume={210},
  number={2},
  pages={208--216},
  year={2007},
  publisher={Company of Biologists}
}

@article{sak2014long,
  title={Long short-term memory based recurrent neural network architectures for large vocabulary speech recognition},
  author={Sak, Ha{\c{s}}im and Senior, Andrew and Beaufays, Fran{\c{c}}oise},
  journal={arXiv preprint arXiv:1402.1128},
  year={2014}
}

@article{alqaham202516,
  title={16 Ways to Gallop: Energetics and Body Dynamics of High-Speed Quadrupedal Gaits},
  author={Alqaham, Yasser G and Cheng, Jing and Gan, Zhenyu},
  journal={arXiv preprint arXiv:2503.13716},
  year={2025}
}

@article{alqaham2024sixteen,
  title={16 Ways to Gallop: Energetics and Body Dynamics of High-Speed Quadrupedal Gaits},
  author={Alqaham, Ahmed and Gan, Zhenyu},
  journal={IEEE Transactions on Robotics},
  year={2024},
  note={Early Access},
  doi={10.1109/TRO.2024.3380201}
}

@article{starkey1989harnessing,
  title={Harnessing and implements for animal traction},
  author={Starkey, Paul},
  journal={A Publication of the Deutsches Zentrum f{\"u}r Entwicklungstechnologien--GATE},
  year={1989}
}

@article{sandberg2020review,
  title={Review of kinematic analysis in dogs},
  author={Sandberg, Gabriella S and Torres, Bryan T and Budsberg, Steven C},
  journal={Veterinary Surgery},
  volume={49},
  number={6},
  pages={1088--1098},
  year={2020},
  publisher={Wiley Online Library}
}

@article{wilshin2020dog,
  title={Dog galloping on rough terrain exhibits similar limb co-ordination patterns and gait variability to that on flat terrain},
  author={Wilshin, Simon and Reeve, Michelle A and Spence, Andrew J},
  journal={Bioinspiration \& Biomimetics},
  volume={16},
  number={1},
  pages={015001},
  year={2020},
  publisher={IOP Publishing}
}

@article{serra2020improving,
  title={Improving gait classification in horses by using inertial measurement unit (IMU) generated data and machine learning},
  author={Serra Bragan{\c{c}}a, FM and Broom{\'e}, S and Rhodin, Marie and Bj{\"o}rnsd{\'o}ttir, S and Gunnarsson, V and Voskamp, JP and Persson-Sjodin, E and Back, W and Lindgren, Gabriella and Novoa-Bravo, M and others},
  journal={Scientific reports},
  volume={10},
  number={1},
  pages={17785},
  year={2020},
  publisher={Nature Publishing Group UK London}
}

@article{Heim2020,
  title = {A little damping goes a long way: a simulation study of how damping influences task-level stability in running},
  volume = {16},
  ISSN = {1744-957X},
  url = {http://dx.doi.org/10.1098/rsbl.2020.0467},
  DOI = {10.1098/rsbl.2020.0467},
  number = {9},
  journal = {Biology Letters},
  publisher = {The Royal Society},
  author = {Heim,  Steve and Millard,  Matthew and Le Mouel,  Charlotte and Badri-Spr\"{o}witz,  Alexander},
  year = {2020},
  month = sep,
  pages = {20200467}
}

@article{sheppard2022stride,
  title={Stride-level analysis of mouse open field behavior using deep-learning-based pose estimation},
  author={Sheppard, Keith and Gardin, Justin and Sabnis, Gautam S and Peer, Asaf and Darrell, Megan and Deats, Sean and Geuther, Brian and Lutz, Cathleen M and Kumar, Vivek},
  journal={Cell reports},
  volume={38},
  number={2},
  year={2022},
  publisher={Elsevier}
}

@article{thorsrud2021description,
  title={Description of breed ancestry and genetic health traits in arctic sled dog breeds},
  author={Thorsrud, Joseph A and Huson, Heather J},
  journal={Canine medicine and genetics},
  volume={8},
  pages={1--13},
  year={2021},
  publisher={Springer}
}

@inproceedings{lawrence1993experimental,
  title={Experimental methods in draught animal research},
  author={Lawrence, Peter R and Pearson, R Anne},
  booktitle={Research for Development of Animal Traction in West Africa.(Eds PR Lawrence, K. Lawrence, JT Dijkman and PH Starkey). Proceedings of the Fourth Workshop of the West Africa Animal Traction Network},
  volume={9},
  pages={187--198},
  year={1993}
}

@inproceedings{starkey1989animal,
  title={Animal-Drawn implements: An overview of recent research and development},
  author={Starkey, Paul and Sims, Brian},
  booktitle={ACIAR proceedings},
  number={27},
  pages={248--257},
  year={1989}
}

@article{geyer2010muscle,
  title={A muscle-reflex model that encodes principles of legged mechanics produces human walking dynamics and muscle activities},
  author={Geyer, Hartmut and Herr, Hugh},
  journal={IEEE Transactions on Neural Systems and Rehabilitation Engineering},
  volume={18},
  number={3},
  pages={263--273},
  year={2010},
  publisher={IEEE},
  doi={10.1109/TNSRE.2010.2047592}
}

@article{full1999templates,
  title={Templates and anchors: neuromechanical hypotheses of legged locomotion on land},
  author={Full, Robert J and Koditschek, Daniel E},
  journal={Journal of Experimental Biology},
  volume={202},
  number={23},
  pages={3325--3332},
  year={1999},
  publisher={The Company of Biologists},
  doi={10.1242/jeb.202.23.3325}
}

@article{biewener2003animal,
  title={Animal locomotion},
  author={Biewener, Andrew A},
  journal={Current biology},
  volume={13},
  number={18},
  pages={R749--R752},
  year={2003},
  publisher={Elsevier}
}

@article{dickinson2000animal,
  title={How animals move: An integrative view},
  author={Dickinson, MH and Farley, CT and Full, RJ and Koehl, MAR and Kram, R and Lehman, S},
  journal={Science},
  volume={288},
  number={5463},
  pages={100--106},
  year={2000},
  publisher={American Association for the Advancement of Science}
}

@article{biancardi2012biomechanical,
  title={Biomechanical determinants of transverse and rotary gallop in cursorial mammals},
  author={Biancardi, Gabriele and Minetti, Alberto E},
  journal={Journal of experimental biology},
  volume={215},
  number={1},
  pages={185--192},
  year={2012},
  publisher={Company of Biologists}
}

@article{geyer2006spring,
  title={Spring-mass model: predicting force, step frequency, center-of-mass trajectory and metabolic cost during running in different environments},
  author={Geyer, Hartmut and Seyfarth, Roman and Blickhan, Reinhard},
  journal={Journal of biomechanics},
  volume={39},
  number={10},
  pages={1929--1941},
  year={2006},
  publisher={Elsevier}
}

@article{fukuhara2018spontaneous,
  title={Spontaneous gait transition to high-speed galloping by reconciliation between body support and propulsion},
  author={Fukuhara, Akira and Owaki, Dai and Kano, Takeshi and Kobayashi, Ryo and Ishiguro, Akio},
  journal={Advanced robotics},
  volume={32},
  number={15},
  pages={794--808},
  year={2018},
  publisher={Taylor \& Francis}
}

@article{Ding2024RALSymmetry,
  title = {Breaking Symmetries Leads to Diverse Quadrupedal Gaits},
  volume = {9},
  ISSN = {2377-3774},
  url = {http://dx.doi.org/10.1109/LRA.2024.3384908},
  DOI = {10.1109/lra.2024.3384908},
  number = {5},
  journal = {IEEE Robotics and Automation Letters},
  publisher = {Institute of Electrical and Electronics Engineers (IEEE)},
  author = {Ding,  Jiayu and Gan,  Zhenyu},
  year = {2024},
  month = may,
  pages = {4782–4789}
}

@article{Hildebrand1977,
  doi = {10.2307/1379571},
  url = {https://doi.org/10.2307/1379571},
  year = {1977},
  month = may,
  publisher = {Oxford University Press ({OUP})},
  volume = {58},
  number = {2},
  pages = {131--156},
  author = {M. Hildebrand},
  title = {Analysis of Asymmetrical Gaits},
  journal = {Journal of Mammalogy}
}

@article{Hilderbrand1989Quadrupedal,
  doi = {10.2307/1311182},
  url = {https://doi.org/10.2307/1311182},
  year = {1989},
  month = dec,
  publisher = {Oxford University Press ({OUP})},
  volume = {39},
  number = {11},
  pages = {766--775},
  author = {Milton Hildebrand},
  title = {The Quadrupedal Gaits of Vertebrates},
  journal = {{BioScience}}
}

@article{Hoyt1981gaitenergetic,
  doi = {10.1038/292239a0},
  url = {https://doi.org/10.1038/292239a0},
  year = {1981},
  month = jul,
  publisher = {Springer Science and Business Media {LLC}},
  volume = {292},
  number = {5820},
  pages = {239--240},
  author = {Donald F. Hoyt and C. Richard Taylor},
  title = {Gait and the energetics of locomotion in horses},
  journal = {Nature}
}

@article{Golubitsky1999,
  title = {Symmetry in locomotor central pattern generators and animal gaits},
  volume = {401},
  ISSN = {1476-4687},
  url = {http://dx.doi.org/10.1038/44416},
  DOI = {10.1038/44416},
  number = {6754},
  journal = {Nature},
  publisher = {Springer Science and Business Media LLC},
  author = {Golubitsky,  Martin and Stewart,  Ian and Buono,  Pietro-Luciano and Collins,  J. J.},
  year = {1999},
  month = oct,
  pages = {693–695}
}

@inproceedings{ames2005sufficient,
  title={Sufficient conditions for the existence of Zeno behavior},
  author={Ames, Aaron D and Abate, Alessandro and Sastry, Shankar},
  booktitle={Proceedings of the 44th IEEE Conference on Decision and Control},
  pages={696--701},
  year={2005},
  organization={IEEE}
}

@inproceedings{pace2017piecewise,
  title={Piecewise-differentiable trajectory outcomes in mechanical systems subject to unilateral constraints},
  author={Pace, Andrew M and Burden, Samuel A},
  booktitle={Proceedings of the 20th International Conference on Hybrid Systems: Computation and Control},
  pages={243--252},
  year={2017}
}

@book{westervelt2018feedback,
  title={Feedback control of dynamic bipedal robot locomotion},
  author={Westervelt, Eric R and Grizzle, Jessy W and Chevallereau, Christine and Choi, Jun Ho and Morris, Benjamin},
  year={2018},
  publisher={CRC press}
}

@article{GanAllCommonBipedal,
  doi = {10.1098/rsif.2018.0455},
  url = {https://doi.org/10.1098/rsif.2018.0455},
  year = {2018},
  month = sep,
  publisher = {The Royal Society},
  volume = {15},
  number = {146},
  pages = {20180455},
  author = {Zhenyu Gan and Yevgeniy Yesilevskiy and Petr Zaytsev and C. David Remy},
  title = {All common bipedal gaits emerge from a single passive model},
  journal = {Journal of The Royal Society Interface}
}

@ARTICLE{GanDynamicSimilarity,
  author={Gan, Zhenyu and Jiao, Ziyuan and Remy, C. David},
  journal={IEEE Robotics and Automation Letters}, 
  title={On the Dynamic Similarity Between Bipeds and Quadrupeds: A Case Study on Bounding}, 
  year={2018},
  volume={3},
  number={4},
  pages={3614-3621},
  doi={10.1109/LRA.2018.2854923}}

@article{DingJerboa2022,
  doi = {10.3389/fbioe.2022.804826},
  url = {https://doi.org/10.3389/fbioe.2022.804826},
  year = {2022},
  month = apr,
  publisher = {Frontiers Media {SA}},
  volume = {10},
  author = {Jiayu Ding and Talia Y. Moore and Zhenyu Gan},
  title = {A Template Model Explains Jerboa Gait Transitions Across a Broad Range of Speeds},
  journal = {Frontiers in Bioengineering and Biotechnology}
}

@article{Farley1991trigger,
  doi = {10.1126/science.1857965},
  url = {https://doi.org/10.1126/science.1857965},
  year = {1991},
  month = jul,
  publisher = {American Association for the Advancement of Science ({AAAS})},
  volume = {253},
  number = {5017},
  pages = {306--308},
  author = {C. Farley and C. Taylor},
  title = {A mechanical trigger for the trot-gallop transition in horses},
  journal = {Science}
}

@article{Rooney1985,
  title = {The mechanics of horses pulling loads},
  volume = {5},
  ISSN = {0737-0806},
  url = {http://dx.doi.org/10.1016/S0737-0806(85)80010-1},
  DOI = {10.1016/s0737-0806(85)80010-1},
  number = {6},
  journal = {Journal of Equine Veterinary Science},
  publisher = {Elsevier BV},
  author = {Rooney,  James R. and Turner,  Larry W.},
  year = {1985},
  month = jan,
  pages = {355–359}
}

@article{Bukhari2023,
  title = {Assessing the impact of draught load pulling on welfare in equids},
  volume = {10},
  ISSN = {2297-1769},
  url = {http://dx.doi.org/10.3389/fvets.2023.1214015},
  DOI = {10.3389/fvets.2023.1214015},
  journal = {Frontiers in Veterinary Science},
  publisher = {Frontiers Media SA},
  author = {Bukhari,  Syed S. U. H. and Parkes,  Rebecca S. V.},
  year = {2023},
  month = aug 
}

@misc{RobotLoadPathPlaning2024,
  doi = {10.48550/ARXIV.2404.12220},
  url = {https://arxiv.org/abs/2404.12220},
  author = {Zhang,  Wentao and Xu,  Shaohang and Zuo,  Gewei and Zhu,  Lijun},
  title = {Hybrid Dynamics Modeling and Trajectory Planning for a Cable-Trailer System with a Quadruped Robot},
  publisher = {arXiv},
  year = {2024},
  copyright = {arXiv.org perpetual,  non-exclusive license}
}

\end{document}